\documentclass[%
preprint,
showpacs,
nofootinbib,
 amsmath,amssymb,
 aps,
prd,
]{revtex4-1}
\usepackage{slashed}
\usepackage{graphicx}
\usepackage{dcolumn}
\usepackage{bm}

\begin{document}


\title{Metastable dark energy}
\author{Ricardo G. Landim}
\email{rlandim@if.usp.br}
 \affiliation{%
 Instituto de F\'isica, Universidade de S\~ao Paulo\\
 Caixa Postal 66318,  05314-970 S\~ao Paulo, S\~ao Paulo, Brazil
}%


\author{Elcio Abdalla}
\email{eabdalla@usp.br}
\affiliation{%
 Instituto de F\'isica, Universidade de S\~ao Paulo\\
 Caixa Postal 66318,  05314-970 S\~ao Paulo, S\~ao Paulo, Brazil
}%

\date{\today}

\begin{abstract}

We build a model of metastable dark energy, in which the observed vacuum energy is the value of the scalar potential at the false vacuum. The scalar potential is given by a sum of even self-interactions up to order six. The deviation from the Minkowski vacuum is due to a term suppressed by the Planck scale. The decay time of the metastable vacuum can easily accommodate a mean life time compatible with the age of the universe. The metastable dark energy is also embedded into a model with $SU(2)_R$ symmetry. The dark energy doublet and the dark matter doublet naturally interact with each other. A three-body decay of the dark energy particle into (cold and warm) dark matter  can be as long as large fraction of the age of the universe, if the mediator is  massive enough, the lower bound being at intermediate energy level some orders below the grand unification scale. Such a decay shows a different form of interaction between dark matter and dark energy, and the model opens a new window to investigate the dark sector from the point-of-view of particle physics.


\end{abstract}

\maketitle



\section{Introduction}

At the present age, around ninety five percent of the universe  corresponds to two kinds of energy whose nature is largely unknown. The first one, named dark energy,  is believed to be responsible for the current accelerated expansion of the universe \cite{reiss1998, perlmutter1999} and  is  dominant at present time ($\sim$ 68\%) \cite{Planck2013cosmological}.  In addition to the baryonic matter (5\%), the remaining $27\%$ of the energy content of the universe is a form of matter that interacts, in principle, only gravitationally, known as dark matter. The simplest dark energy candidate is the cosmological constant, whose equation of state is in agreement with the Planck results \cite{Planck2013cosmological}.

This attempt, however, suffers from the so-called cosmological constant problem, a huge discrepancy of 120 orders of magnitude between the theoretical (though rather speculative) prediction and the observed data \cite{Weinberg:1988cp}. Such a huge disparity motivates physicists to look into more sophisticated models. This can be done either looking for a deeper understanding of where the cosmological constant comes from, if one wants to derive it from first principles, or considering other possibilities for accelerated expansion, such as  modifications of general relativity (GR), additional matter fields and so on (see \cite{copeland2006dynamics, dvali2000, yin2005} and references therein).  Moreover, the theoretical origin of this constant is still an open question, with several attempts but with no definitive answer yet.

There is a wide range of alternatives to the cosmological constant, which includes canonical and non-canonical scalar fields   \cite{peebles1988,ratra1988,Frieman1992,Frieman1995,Caldwell:1997ii,Padmanabhan:2002cp,Bagla:2002yn,ArmendarizPicon:2000dh,Brax1999,Copeland2000,Landim:2015upa,micheletti2009}, vector fields \cite{Koivisto:2008xf,Bamba:2008ja,Emelyanov:2011ze,Emelyanov:2011wn,Emelyanov:2011kn,Kouwn:2015cdw,Landim:2016dxh,costa2014}, holographic dark energy \cite{Hsu:2004ri,Li:2004rb,Pavon:2005yx,Wang:2005jx,Wang:2005pk,Wang:2005ph,Wang:2007ak,Landim:2015hqa}, modifications of gravity and different kinds of cosmological fluids \cite{copeland2006dynamics, dvali2000, yin2005,Dymnikova:2001ga,Dymnikova:2001jy,Mukhopadhyay:2007ed}. 

In addition, the two components of the dark sector may interact with each other \cite{Wetterich:1994bg,Amendola:1999er} (see \cite{Wang:2016lxa} for a recent review), since their densities are comparable and the interaction can eventually alleviate the coincidence problem \cite{Zimdahl:2001ar,Chimento:2003iea}. Phenomenological models have been widely explored in the literature \cite{Amendola:1999er,Guo:2004vg,Cai:2004dk,Guo:2004xx,Bi:2004ns,Gumjudpai:2005ry,yin2005,Wang:2005jx,Wang:2005pk,Wang:2005ph,Wang:2007ak,Costa:2013sva,Abdalla:2014cla,Costa:2014pba,Costa:2016tpb,Marcondes:2016reb}. On the other hand, field theory models that aim a consistent description of the dark energy/dark matter interaction are still few \cite{Farrar:2003uw,Abdalla:2012ug,D'Amico:2016kqm,micheletti2009}. 

Here we propose a model of metastable dark energy, in which the dark energy is a scalar field with a potential given by the sum of even self-interactions up to order six. The parameters of the model can be adjusted in such a way that the difference between the energy of the true vacuum and the energy of the false one is the observed vacuum energy  ($10^{-47}$ GeV$^4$). Other models of false vacuum decay were proposed in \cite{Stojkovic:2007dw,Greenwood:2008qp,Abdalla:2012ug} with different potentials. A different mechanism of metastable dark energy (although with same name) is presented in \cite{Shafieloo:2016bpk}. Furthermore, a dark $SU(2)_R$ model is presented, where the dark energy doublet and the dark matter doublet naturally interact with each other. Such an interaction opens a new window to investigate the dark sector from the point-of-view of particle physics. Models with $SU(2)_R$ symmetry are well-known in the literature as extensions of the standard model  introducing the so-called left-right symmetric models \cite{Aulakh:1998nn,Duka:1999uc,Dobrescu:2015qna,Dobrescu:2015jvn,Ko:2015uma}. Recently,  dark matter has also been taken into account  \cite{Bezrukov:2009th,Esteves:2011gk,An:2011uq,Nemevsek:2012cd,Bhattacharya:2013nya,Heeck:2015qra,Garcia-Cely:2015quu,Berlin:2016eem}.  However, there is no similar effort to insert dark energy in a model of particle physics. We begin to attack this issue in this paper, with the dark $SU(2)_R$ model.

The remainder of this paper is structured as follows. In Sect. \ref{MSDE} we present a model of metastable dark energy. It is embedded into a dark $SU(2)_R$ model in Sect. \ref{darSU2} and we summarize our results in Sect. \ref{concluSU2}. We use natural units ($\hbar=c=1$) throughout the text.

\section{A model of metastable dark energy}\label{MSDE}

The current stage of accelerated expansion of the universe will be described by a canonical scalar field $\varphi$ at a local minimum $\varphi_0$ of its potential $V(\varphi)$, while the true minimum of $V(\varphi)$ is at $\varphi_{\pm}=\langle\varphi\rangle$. The energy of the true vacuum is below the zero energy of the false vacuum, so that this difference is interpreted as the observed value of the vacuum energy ($10^{-47}$ GeV$^4$). 

We assume that by some mechanism the scalar potential is positive definite (as e.g. in supersymmetric models) and the true vacuum lies at zero energy. As we will see below this value is adjusted by the mass of the scalar field and the coefficient of the quartic and sixth-order interaction. The rate at which the false vacuum decays into the true vacuum state will be calculated.

The process of barrier penetration in which the metastable false vacuum decays into the stable true vacuum is similar to the  old inflationary scenario and it occurs through the formation of bubbles of true vacuum in a false vacuum background. After the barrier penetration the bubbles grow at the speed of light and eventually collide with other bubbles until all  space is in the lowest energy state. The energy release in the process can produce new particles and  a Yukawa interaction $g \varphi\bar{\psi}\psi$ can account for the production of a fermionic field which can be the pressureless fermionic dark matter.  However, as we will see, the vacuum time decay is of the order of the age of the universe, so another dominant process for the production of cold dark matter should be invoked in order to recapture the standard cosmology.

If one considers a scalar field $\varphi$ with the even self-interactions up to order six, one gets

\begin{equation}\label{VScalar}
 V(\varphi)=\frac{m^2}{2}\varphi^2-\frac{\lambda}{4}\varphi^4+\frac{\lambda^2}{32 m^2}\varphi^6\quad,
\end{equation} 

\noindent where $m$ and $\lambda$ are positive free parameters of the theory and the coefficient of the $\varphi^6$ interaction is chosen in such a way that the potential (\ref{VScalar}) is a perfect square. This choice will be useful to calculate the false vacuum decay rate.

The potential (\ref{VScalar}) has  extrema  at $\varphi_0=0$, $\varphi_{\pm}=\pm\frac{ 2m}{\sqrt{\lambda}}$ and $\varphi_1=\frac{\varphi_{\pm}}{\sqrt{3}}$, but it is zero in all of the minima ($\varphi_0$ and $\varphi_{\pm}$). In order to have a cosmological constant, the potential should  deviate slightly from the perfect square (\ref{VScalar}).  Once the coupling present in GR is the Planck mass $M_{pl}$ it is natural to expect that the deviation from the Minkowski vacuum is due to a term proportional to $M^{-2}_{pl}$. Thus we assume that the potential (\ref{VScalar}) has a small deviation given by $\frac{\varphi^6}{M_{pl}^2}$. Although the value of the scalar field at the minimum point $\varphi_\pm$ also changes, the change is very small and we can consider that the scalar field at the true vacuum is still $\pm\frac{ 2m}{\sqrt{\lambda}}$. The difference between the true vacuum and the false one is 
\begin{equation}
V(\varphi_0)-V(\varphi_\pm)\approx\frac{64m^6}{\lambda^3M_{pl}^2}\quad.
\label{eq:cosmoconst}
\end{equation}

As usual in quantum field theory it is expected that the parameter $\lambda$ is smaller than one, thus, if we assume $\lambda \sim 10^{-1}$, the Eq. (\ref{eq:cosmoconst}) gives $\sim 10^{-47}$ GeV$^4$ for $m\sim \mathcal {O}(\text{MeV})$. Bigger values of $\lambda$ imply smaller values of $m$. Therefore, the cosmological constant is determined by the mass parameter and the coupling of the quartic interaction.

\begin{figure}%
\includegraphics[scale=0.7]{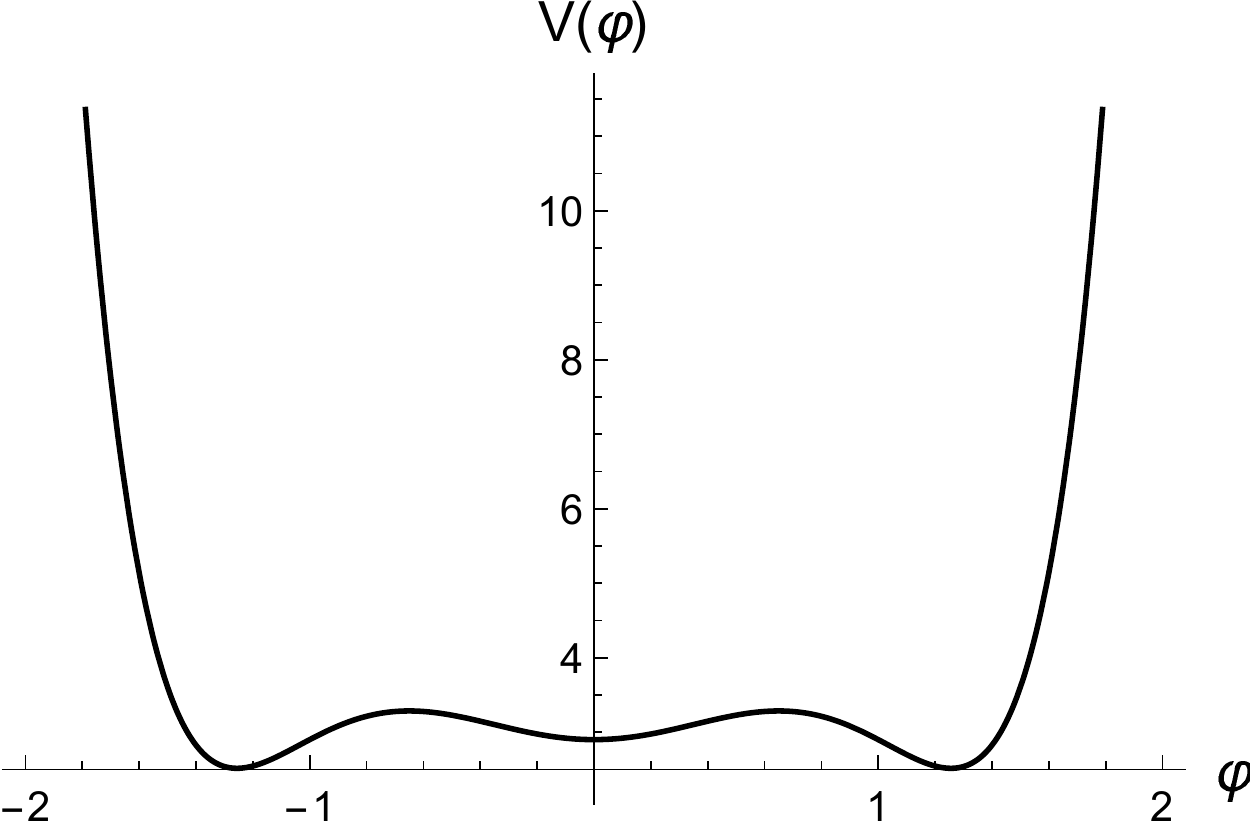}%
\caption{Scalar potential (\ref{VScalar}) with arbitrary parameters and values. The difference between the true vacuum at $\varphi_\pm\approx \pm 1.2 $   and  the false vacuum at $\varphi_0=0$ is $\sim 10^{-47}$ GeV$^4$.}%
\label{potentialFig}%
\end{figure}

 The potential (\ref{VScalar}) with the term $\frac{\varphi^6}{M_{pl}^2}$ is shown in Figure \ref{potentialFig}.

\subsection{Decay rate}\label{sec:Decay}

The computation of the decay rate is based on the semi-classical theory presented in \cite{Coleman:1977py}. The energy of the false vacuum state at which $\langle \varphi\rangle=0$ is given by \cite{Weinberg:1996kr}
\begin{equation}
E_0=-\lim_{T\rightarrow\infty}\frac{1}{T}\ln\left[\int{\exp\left(-S_E[\varphi;T]\right)\prod_{\vec{x},t}\,d\varphi(\vec{x},t)}\right]\quad ,
\label{eq:E0}
\end{equation}
where $S_E[\varphi;T]$ is the Euclidean action,
\begin{equation}
S_E=\int\,d^3x\int_{-\frac{T}{2}}^{+\frac{T}{2}}\,dt\left[\frac{1}{2}\left(\frac{\partial\varphi}{\partial t}\right)^2+\frac{1}{2}\left(\nabla\varphi\right)^2+V(\varphi)\right]\quad.
\label{eq:S}
\end{equation}

 The imaginary part of $E_0$ gives the decay rate and all the fields $\varphi(\vec{x},t)$ integrated in Eq. (\ref{eq:E0})  satisfy the boundary conditions 
\begin{equation}\label{boundarycond}\varphi(\vec{x},+T/2)=\varphi(\vec{x},-T/2)=0\quad .
\end{equation} 

The action (\ref{eq:S}) is stationary under variation of the fields that satisfy the equations
\begin{equation}
\frac{\delta S_E}{\delta\varphi}=-\frac{\partial^2\varphi}{\partial t^2}-\nabla^2\varphi+V'(\varphi)=0
\label{eq:Seq}
\end{equation}
and  are subject to the boundary conditions (\ref{boundarycond}). In order to get the solution of Eq. (\ref{eq:Seq}) we make an ansatz that the field $\varphi(\vec{x},t)$ is invariant under rotations around $\vec{x}_0,t_0$ in four dimensions, which in turn is valid for large $T$ \cite{Coleman:1977th}. The ansatz is 
\begin{equation}
\varphi(\vec{x},t)=\varphi(\rho) \quad \text{with } \quad \rho\equiv\sqrt{(\vec{x}-\vec{x}_0)^2+(t-t_0)^2}\quad .
\label{eq:ansatz}
\end{equation}

In terms of the Eq. (\ref{eq:ansatz}), the field equations (\ref{eq:Seq}) becomes
\begin{equation}
\frac{d^2\varphi}{d \rho^2}+\frac{3}{\rho}\frac{d\varphi}{d\rho}=V'(\varphi)\quad .
\label{eq:Seq2}
\end{equation}

The above equation of motion  is analogous to that of a particle at position $\varphi$  moving in a time $\rho$, under the influence of a potential $-V(\varphi)$ and a viscous force $-\frac{3}{\rho}\frac{d\varphi}{d\rho}$. This particle travels from an initial value $\varphi_i$ and $\rho=0$ and reaches $\varphi=0$ at $\rho\rightarrow\infty$. The Euclidean action (\ref{eq:S}) for the rotation invariant solution becomes
\begin{equation}
S_E=\int_{0}^{\infty}2\pi^2 \rho^3\,d\rho\left[\frac{1}{2}\left(\frac{\partial\varphi}{\partial \rho}\right)^2+V(\varphi)\right]\quad .
\label{eq:Srot}
\end{equation}

The metastable vacuum decay into the true vacuum is seen as the formation of bubbles of true vacuum surrounded by the false vacuum outside. The friction term $\frac{d\varphi}{d\rho}$ is different from zero only at the bubble wall, since the field is at rest inside and outside. The decay rate per volume of the false vacuum, in the semi-classical approach, is of order 
\begin{equation}
\frac{\Gamma}{V}\approx M^{-4}\exp(-S_E)\quad ,
\label{eq:decayrate}
\end{equation}
where $M$ is some mass scale. When $S_E$ is large the barrier penetration is suppressed and the mass scale is not important. This is the case when the energy of the true vacuum is slightly below the energy of the false vacuum, by an amount $\epsilon$, considered here as small as $\epsilon\sim 10^{-47}$ GeV$^4$. On the other hand, the potential $V(\varphi)$ is not small between $\varphi_0$ and $\varphi_\pm$.

We will use the so-called `thin wall approximation', in which $\varphi$ is taken to be inside of a four-dimensional sphere of large radius $R$. For a thin wall we can consider $\rho\approx R$  in this region and since $R$ is large we can neglect the viscous term, which is proportional to $3/R$ at the wall. The action (\ref{eq:Srot}) in this approximation is 
\begin{equation}
S_E\simeq -\frac{\pi^2}{2}R^4\epsilon+2\pi^2R^3S_1\quad ,
\label{eq:Sthin}
\end{equation}
where $S_1$ is a surface tension, given by
\begin{equation}
S_1= \sqrt{2}\int_{\varphi_0}^{\varphi_+}\,d\varphi{ \sqrt{V}}\quad ,
\label{eq:S1}
\end{equation}
for small $\epsilon$.  The action (\ref{eq:Sthin}) is stationary at the radius 
\begin{equation}
R\simeq \frac{3S_1}{\epsilon}\quad ,
\label{eq:Rmin}
\end{equation}
and at the stationary point the action (\ref{eq:Sthin}) becomes
\begin{equation}
S_E\simeq \frac{27\pi^2S_1^4}{2\epsilon^3}\quad .
\label{eq:SthinA}
\end{equation}

Using the potential (\ref{VScalar}) into Eq. (\ref{eq:S1}) we obtain\footnote{The term $\frac{\varphi^6}{M^2_{pl}}$ is very small and can be ignored. }
\begin{equation}
S_1= \frac{m^3}{\lambda}\quad ,
\label{eq:S1Pot}
\end{equation}
which in turn gives the Euclidean action at the stationary point in the thin wall approximation (\ref{eq:Sthin}) 
\begin{equation}
S_E\simeq \frac{27\pi^2m^{12}}{2\lambda^4\epsilon^3}\quad.
\label{eq:SthinPot}
\end{equation}

Substituting the action (\ref{eq:SthinPot}) into the decay rate (\ref{eq:decayrate}) with $\epsilon\sim 10^{-47}$ GeV$^4$ and the mass scale being $M\sim 1$ GeV for simplicity, we have
\begin{equation}
\frac{\Gamma}{V}\approx \exp\left[-10^{143}\left(\frac{m}{\text{GeV}}\right)^{12}\lambda^{-4}\right] \text{GeV}^4
\quad .
\label{eq:decayrate2}
\end{equation}

The decay time is obtained inverting the above expression,
\begin{equation}
t_{decay}\approx 10^{-25}\left\{\exp\left[10^{143}\left(\frac{m}{\text{GeV}}\right)^{12}\lambda^{-4}\right] \right\}^{1/4}\text{s}\quad.
\label{eq:tdecayrate}
\end{equation}

The expression for the decay time gives the lowest value of the mass parameter $m$ for which (\ref{eq:tdecayrate}) has at least the age of the universe ($10^{17}$ s). Therefore the mass parameter should be
\begin{equation}
m\gtrsim 10^{-12} \text{GeV}\quad ,
\label{eq:m}
\end{equation}
for $\lambda\sim 10^{-1}$. Thus, it is in agreement with  the values for $m$ at which the scalar potential describes the observed vacuum energy, as discussed in the last section.  The mass of the scalar field can be smaller if the coupling $\lambda$ is also smaller than $10^{-1}$. The decay rate (\ref{eq:decayrate2}) is strongly suppressed for larger values of $m$. The bubble radius given in Eq. (\ref{eq:Rmin}) for the mass parameter (\ref{eq:m}) is $R\gtrsim 0.03$ cm.

Notice that the axion would still be a possibility, although it arises in a quite different context. We can also consider the gravitational effect in the computation of the decay rate. In this case the new action $\bar{S}$ has the Einstein-Hilbert term $\frac{M_{pl}^2}{2}\mathcal{R}$, where  $\mathcal{R}$ is the Ricci scalar. The relation between the new action  $\bar{S}$ and the old one $S_E$ can be deduced using the thin wall approximation and it gives \cite{Coleman:1980aw}
\begin{equation}
\bar{S}=\frac{S_E}{\left(1+\left(\frac{R}{2\Delta}\right)^2\right)^2}\quad ,
\label{Sbar}
\end{equation}
where $S_E$ and $R$ are given by Eqs. (\ref{eq:SthinA}) and (\ref{eq:Rmin}), respectively, in the absence of gravity, and $\Delta=\frac{\sqrt{3}M_{pl}}{\sqrt{\epsilon}}$ is the value of the bubble radius when it is equal to the   Schwarzschild radius associated with the energy released by the conversion of false vacuum to true one. 

 For $\epsilon\sim 10^{-47}$ GeV$^4$ we get $\Delta\sim 10^{27}$ cm, thus the gravitational correction $R/\Delta$ is very small. Larger values of $m$ give larger $R$, implying that  the gravitational effect should be taken into account. Even so, the decay rate is still highly suppressed.

\section{A dark $SU(2)_R$ model}\label{darSU2}

As an example of how the metastable dark energy can be embedded into a dark sector model we restrict our attention to  a model with  $SU(2)_R$ symmetry. Both dark energy and dark matter are doublets under $SU(2)_R$ and singlets under any other symmetry. Presumably, the dark sector interacts with the standard model particles only through gravity. After the spontaneous symmetry breaking by the dark Higgs field $\phi$, the gauge bosons $W_d^+$, $W_d^-$ and $Z_d$ acquire the same mass given by $m_W=m_Z=g v/2$, where $v$ is the VEV of the dark Higgs. The dark $SU(2)_R$ model contains a  dark matter candidate $\psi$,  a dark neutrino $\nu_d$ (which can be much lighter than $\psi$), and the dark energy doublet $\varphi$, which contains $\varphi^0$ and $\varphi^+$,  the latter being  the heaviest particle. After  symmetry breaking $\varphi^0$ and $\varphi^+$ have different masses and both have a potential given by Eq. (\ref{VScalar}) plus the deviation $\frac{(\varphi^\dagger\varphi)^3}{M_{pl}^2}$  The interaction between the fields are given by the Lagrangian
\begin{eqnarray}
 \mathcal{L}_{int}=
g\left(W_{d\mu}^+J_{dW}^{+\mu} +W_{d\mu}^-J_{dW}^{-\mu}+Z_{d\mu}^0J_{dZ}^{0\mu}\right)\quad,
\label{eq:Lint}
\end{eqnarray}
where the currents are 
\begin{eqnarray}
&&\quad J_{dW}^{+\mu} =\frac{1}{\sqrt{2}}[\bar{\nu}_{dR}\gamma^\mu \psi_R+i(\varphi^0\partial^\mu\bar{\varphi}^+-\bar{\varphi}^+\partial^\mu\varphi^0)]\quad , \label{eq:JW+}\\
&&\quad J_{dW}^{-\mu} =\frac{1}{\sqrt{2}}[\bar{\psi}_{R}\gamma^\mu \nu_{dR}+i(\varphi^+\partial^\mu\bar{\varphi}^0-\bar{\varphi}^0\partial^\mu\varphi^+)]
\quad ,\label{eq:JW-}\\
J_{dZ}^{0\mu} &=&\frac{1}{2}[\bar{\nu}_{dR}\gamma^\mu \nu_{dR}-\bar{\psi}_{R}\gamma^\mu \psi_{R}+
i(\varphi^+\partial^\mu\bar{\varphi}^+-\bar{\varphi}^+\partial^\mu\varphi^+)-i(\varphi^0\partial^\mu\bar{\varphi}^0-\bar{\varphi}^0\partial^\mu\varphi^0)]\quad .
\label{eq:JZ0}
\end{eqnarray}

 The currents above are very similar to the ones in the electroweak theory. The main differences are that there is no hypercharge due to $U(1)_Y$ and there is a new doublet, given by $\varphi^+$ and $\varphi^0$. 

Among the interactions shown in Eq. (\ref{eq:Lint}), it is of interest to calculate the decay rate due to the process $\varphi^+\rightarrow \varphi^0+\psi+\nu_d$. The three-body decay leads to a cold dark matter particle whose mass can be accommodated to give the correct relic abundance, to a dark neutrino which is a hot/warm dark matter particle, and to a scalar field $\varphi^0$. Similar to the weak interactions, we assume that the energy involved in the decay is much lower than the mass of the gauge fields, thus the propagator of $W$ is proportional to $g^2/m_W^2$ and the currents interact at a point. We can also define 
\begin{equation}
 \frac{g^2}{8m_W^2}\equiv \frac{G_d}{\sqrt{2}}\quad ,
\label{eq:darkcoupling}
\end{equation}
where $G_d$ is the dark coupling.

		\begin{figure}
		\centering\includegraphics[scale=1]{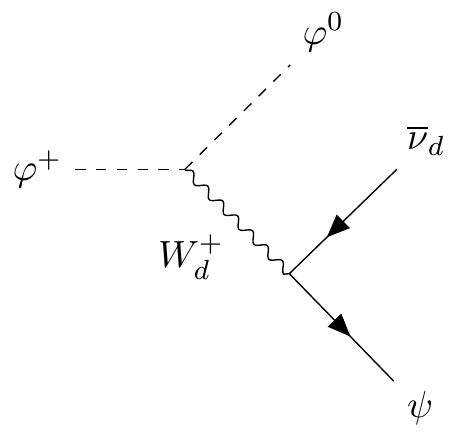}
\caption{Feynman diagram for the decay $\varphi^+\rightarrow \varphi^0+\psi+\nu_d$.}\label{feyndiagram}
\end{figure}

The Feynman diagram for the decay is shown in Fig. \ref{feyndiagram} and the amplitude for the decay is 
\begin{eqnarray}
\mathcal{M}=\frac{g^2}{4m_W^2}(P+p_1)_\mu\overline{u}(p_3)\gamma^\mu(1+\gamma^5)u(p_2)
\quad ,
\label{eq:M}
\end{eqnarray}
where the labels $1$, $2$ and $3$ are used, respectively, for the particles $\varphi^0$, $\psi$ and $\nu_d$. The energy-momentum conservation implies that $P=p_1+p_2+p_3$, where $P$ is the four-momentum of the field $\varphi^+$ and $M$ will be its mass. 

The averaged amplitude squared for the decay $\varphi^+\rightarrow \varphi^0+\psi+\nu_d$ is
\begin{eqnarray}
\overline{ |\mathcal{M}|^2}= 16 G_d^2\left\{2[(P+p_1)\cdot p_2][(P+p_1)\cdot p_3]-(P+p_1)^2(p_2\cdot p_3+m_2m_3)\right\}\quad.
\label{eq:M12}
\end{eqnarray}

Using the energy-momentum conservation  and defining the invariants $s_{ij}$ as $s_{ij}\equiv (p_i+p_j)^2=(P-p_k)^2$, we can reorganize the amplitude squared. The three invariants are not independent, obeying $s_{12}+s_{23}+s_{13}=M^2+m_1^2+m_2^2+m_3^2$ from their definitions and the energy-momentum conservation. With all these steps we  eliminate $s_{13}$ and get
\begin{eqnarray}
\overline{ |\mathcal{M}|^2}= &16 G_d^2[ -2 s_{12}^2-2s_{12}s_{23}+2(M^2+m_1^2+m_2^2+m_3^3)s_{12}+\frac{(m_2+m_3)^2}{2}s_{23}\nonumber\\
&-2m_2m_3(M^2+m_1^2)-2m_1^2M^2-2m_2^2(m_1^2+m_2^2)-\frac{(m_2+m_3)^2}{2} ]\quad .
\label{eq:M2}
\end{eqnarray}

The decay rate can be evaluated from \cite{Agashe:2014kda}
\begin{equation}
d\Gamma= \frac{ 1}{(2\pi)^3}\frac{1}{32M^3}\overline{|\mathcal{M}|^2}d s_{12}d s_{23}\quad , \label{eq:dgamma}
\end{equation}
where for a given value of $s_{12}$, the range of $s_{23}$ is determined by its
values when $\vec{p_2}$ is parallel or antiparallel to $\vec{p_3}$ 
\begin{eqnarray}
(s_{23})_{max}&=& (E_2^*+E_3^*)^2-\left(\sqrt{E_2^{*2}-m_2^2}-\sqrt{E_3^{*2}-m_3^2}\right)^2
\quad ,\label{s23max}\\
(s_{23})_{min}&=& (E_2^*+E_3^*)^2-\left(\sqrt{E_2^{*2}-m_2^2}+\sqrt{E_3^{*2}-m_3^2}\right)^2\quad . 
\label{s23min}
\end{eqnarray}

The energies $E_2^*=(s_{12}	-m_1^2+m_2^2)/(2\sqrt{s_{12}})$ and $E_3^*=(M	-s_{12}-m_3^2)/(2\sqrt{s_{12}})$ are the energies
of particles 2 and 3 in the $s_{12}$ rest frame \cite{Agashe:2014kda}. The invariant $s_{12}$, in turn, has the limits 
\begin{equation}
\label{limits12}
(s_{12})_{max}=(M-m_3)^2, \qquad (s_{12})_{min}=(m_1+m_2)^2\quad . 
\end{equation}

\begin{figure}%
\includegraphics[scale=0.6]{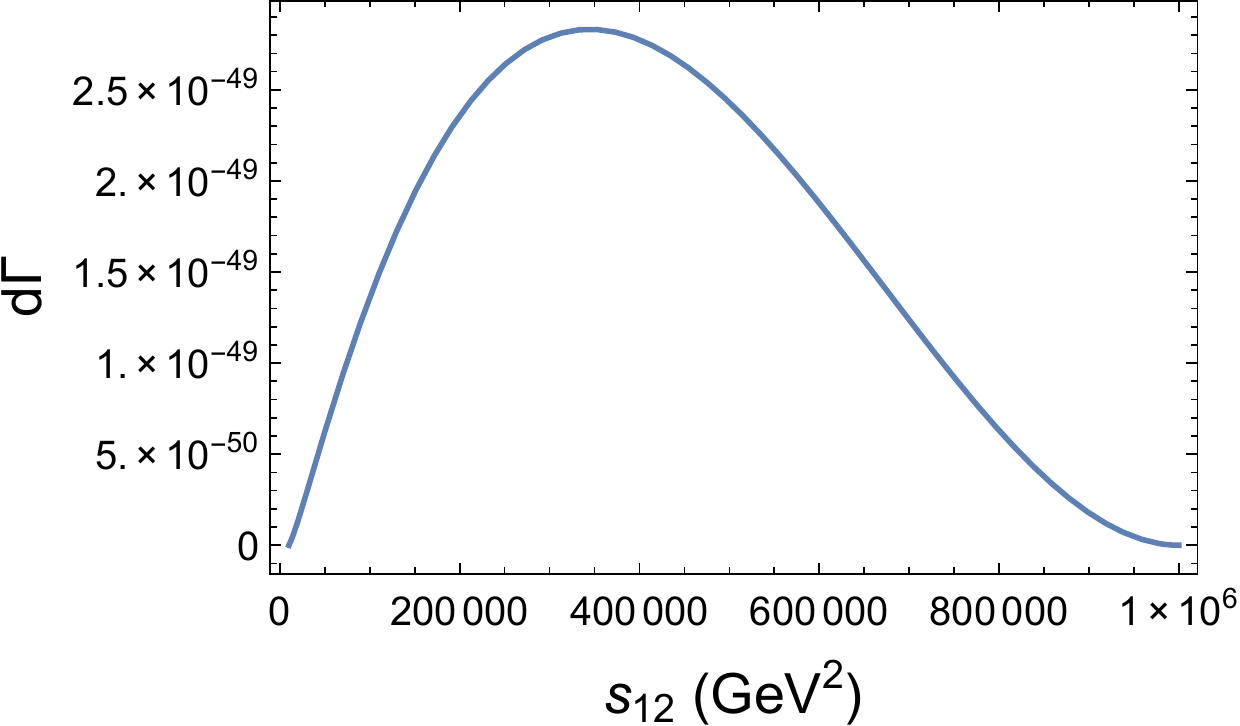}%
\caption{Differential decay rate $d\Gamma$ (\ref{eq:dgamma}) as a function of $s_{12}$ for $M=1000$ GeV, $m_1=1$ MeV, $m_2=100$ GeV, $m_3=0$ and $G_d\sim 10^{-27} $ GeV$^{-2}$.}%
\label{decaywidth}%
\end{figure}

With the limits for $s_{12}$ (\ref{limits12}) and for $s_{23}$ (\ref{s23max})--(\ref{s23min}) and with the amplitude squared  (\ref{eq:M2})  we can integrate Eq. (\ref{eq:dgamma}) for different values of masses, in order to get the decay time $t_{dec}$.

The plot of $d\Gamma$ as a function of $s_{12}$ is shown in Fig. \ref{decaywidth} and the decay rate $\Gamma$ is the area under the curve. For illustrative purposes, we set the mass of the particles as being $M=1000$ GeV, $m_1=1$ MeV, $m_2=100$ GeV and $m_3=0$ GeV. With these values of masses,  the decay time is of the order of the age of the universe ($10^{17}$s) with $G_d\sim 10^{-27}$ GeV$^{-2}$, while  with $G_d\sim 10^{-26}$ GeV$^{-2}$ the decay time is $t_{dec}\sim 10^{15}$s. If $g$ is for instance of the same order of the fine-structure constant,  the gauge bosons $W_d^\pm$ and $Z_d$ have masses  around $10^{11}$ GeV in order to the decay time to be $10^{15}$s. Such decay times are compatible with  phenomenological models of interacting dark energy,  where the coupling is proportional to the Hubble rate  \cite{Costa:2016tpb,Wang:2016lxa}. In addition, depending on the values of the free parameters, the mass of the gauge bosons can be of the same order of the grand unified theories scale.

 \section{Conclusions}\label{concluSU2}

In this paper we presented a model of metastable dark energy, in which the dark energy is a scalar field with a potential given by a sum of even self-interactions up to order six. The parameters of the model can be adjusted in such a way that the difference between the energy of the true vacuum and the energy of the false one is around $10^{-47}$ GeV$^4$. The decay of the false vacuum to the true one is highly suppressed, thus the metastable dark energy can explain the current accelerated expansion of the universe. We do not need a very tiny mass for the scalar field (as it happens for some models of quintessence), in order to get the observed value of the vacuum energy.

The metastable dark energy can be inserted into a more sophisticated model for the dark sector. In this paper we restricted our attention to a Lagrangian invariant under $SU(2)_R$ (before the spontaneous symmetry breaking by the dark Higgs), in which the dark energy doublet and the dark matter doublet naturally interact with each other. The decay of the heaviest particle of the dark energy doublet into the three daughters (dark energy particle, cold and hot dark matter) was calculated and the decay time can be as long as the age of the universe, if the mediator is massive enough. Such a decay shows a different form of interaction between dark matter and dark energy, and the model opens a new window to investigate the dark sector from the point-of-view of particle physics.

\begin{acknowledgments}
 This work is supported by CNPq and FAPESP (Grant No. 2011/18729-1 and 2013/10242-1). 
\end{acknowledgments}

\bibliography{metastableDE}

\providecommand{\noopsort}[1]{}\providecommand{\singleletter}[1]{#1}%
\begin{thebibliography}{77}%
\makeatletter
\providecommand \@ifxundefined [1]{%
 \@ifx{#1\undefined}
}%
\providecommand \@ifnum [1]{%
 \ifnum #1\expandafter \@firstoftwo
 \else \expandafter \@secondoftwo
 \fi
}%
\providecommand \@ifx [1]{%
 \ifx #1\expandafter \@firstoftwo
 \else \expandafter \@secondoftwo
 \fi
}%
\providecommand \natexlab [1]{#1}%
\providecommand \enquote  [1]{``#1''}%
\providecommand \bibnamefont  [1]{#1}%
\providecommand \bibfnamefont [1]{#1}%
\providecommand \citenamefont [1]{#1}%
\providecommand \href@noop [0]{\@secondoftwo}%
\providecommand \href [0]{\begingroup \@sanitize@url \@href}%
\providecommand \@href[1]{\@@startlink{#1}\@@href}%
\providecommand \@@href[1]{\endgroup#1\@@endlink}%
\providecommand \@sanitize@url [0]{\catcode `\\12\catcode `\$12\catcode
  `\&12\catcode `\#12\catcode `\^12\catcode `\_12\catcode `\%12\relax}%
\providecommand \@@startlink[1]{}%
\providecommand \@@endlink[0]{}%
\providecommand \url  [0]{\begingroup\@sanitize@url \@url }%
\providecommand \@url [1]{\endgroup\@href {#1}{\urlprefix }}%
\providecommand \urlprefix  [0]{URL }%
\providecommand \Eprint [0]{\href }%
\providecommand \doibase [0]{http://dx.doi.org/}%
\providecommand \selectlanguage [0]{\@gobble}%
\providecommand \bibinfo  [0]{\@secondoftwo}%
\providecommand \bibfield  [0]{\@secondoftwo}%
\providecommand \translation [1]{[#1]}%
\providecommand \BibitemOpen [0]{}%
\providecommand \bibitemStop [0]{}%
\providecommand \bibitemNoStop [0]{.\EOS\space}%
\providecommand \EOS [0]{\spacefactor3000\relax}%
\providecommand \BibitemShut  [1]{\csname bibitem#1\endcsname}%
\let\auto@bib@innerbib\@empty
\bibitem [{\citenamefont {Riess}\ \emph {et~al.}(1998)\citenamefont {Riess}
  \emph {et~al.}}]{reiss1998}%
  \BibitemOpen
  \bibfield  {author} {\bibinfo {author} {\bibfnamefont {A.~G.}\ \bibnamefont
  {Riess}} \emph {et~al.} (\bibinfo {collaboration} {Supernova Search Team}),\
  }\href {\doibase 10.1086/300499} {\bibfield  {journal} {\bibinfo  {journal}
  {Astron.J.}\ }\textbf {\bibinfo {volume} {116}},\ \bibinfo {pages} {1009}
  (\bibinfo {year} {1998})},\ \Eprint {http://arxiv.org/abs/astro-ph/9805201}
  {arXiv:astro-ph/9805201 [astro-ph]} \BibitemShut {NoStop}%
\bibitem [{\citenamefont {Perlmutter}\ \emph {et~al.}(1999)\citenamefont
  {Perlmutter} \emph {et~al.}}]{perlmutter1999}%
  \BibitemOpen
  \bibfield  {author} {\bibinfo {author} {\bibfnamefont {S.}~\bibnamefont
  {Perlmutter}} \emph {et~al.} (\bibinfo {collaboration} {Supernova Cosmology
  Project}),\ }\href {\doibase 10.1086/307221} {\bibfield  {journal} {\bibinfo
  {journal} {Astrophys.J.}\ }\textbf {\bibinfo {volume} {517}},\ \bibinfo
  {pages} {565} (\bibinfo {year} {1999})},\ \Eprint
  {http://arxiv.org/abs/astro-ph/9812133} {arXiv:astro-ph/9812133 [astro-ph]}
  \BibitemShut {NoStop}%
\bibitem [{\citenamefont {Ade}\ \emph {et~al.}(2014)\citenamefont {Ade} \emph
  {et~al.}}]{Planck2013cosmological}%
  \BibitemOpen
  \bibfield  {author} {\bibinfo {author} {\bibfnamefont {P.~A.~R.}\
  \bibnamefont {Ade}} \emph {et~al.} (\bibinfo {collaboration} {Planck}),\
  }\href {\doibase 10.1051/0004-6361/201321591} {\bibfield  {journal} {\bibinfo
   {journal} {Astron.Astrophys.}\ }\textbf {\bibinfo {volume} {571}},\ \bibinfo
  {pages} {A16} (\bibinfo {year} {2014})},\ \Eprint
  {http://arxiv.org/abs/1303.5076} {arXiv:1303.5076 [astro-ph.CO]} \BibitemShut
  {NoStop}%
\bibitem [{\citenamefont {Weinberg}(1989)}]{Weinberg:1988cp}%
  \BibitemOpen
  \bibfield  {author} {\bibinfo {author} {\bibfnamefont {S.}~\bibnamefont
  {Weinberg}},\ }\href {\doibase 10.1103/RevModPhys.61.1} {\bibfield  {journal}
  {\bibinfo  {journal} {Rev. Mod. Phys.}\ }\textbf {\bibinfo {volume} {61}},\
  \bibinfo {pages} {1} (\bibinfo {year} {1989})}\BibitemShut {NoStop}%
\bibitem [{\citenamefont {Copeland}\ \emph {et~al.}(2006)\citenamefont
  {Copeland}, \citenamefont {Sami},\ and\ \citenamefont
  {Tsujikawa}}]{copeland2006dynamics}%
  \BibitemOpen
  \bibfield  {author} {\bibinfo {author} {\bibfnamefont {E.~J.}\ \bibnamefont
  {Copeland}}, \bibinfo {author} {\bibfnamefont {M.}~\bibnamefont {Sami}}, \
  and\ \bibinfo {author} {\bibfnamefont {S.}~\bibnamefont {Tsujikawa}},\ }\href
  {\doibase 10.1142/S021827180600942X} {\bibfield  {journal} {\bibinfo
  {journal} {Int. J. Mod. Phys.}\ }\textbf {\bibinfo {volume} {D15}},\ \bibinfo
  {pages} {1753} (\bibinfo {year} {2006})},\ \Eprint
  {http://arxiv.org/abs/hep-th/0603057} {arXiv:hep-th/0603057 [hep-th]}
  \BibitemShut {NoStop}%
\bibitem [{\citenamefont {Dvali}\ \emph {et~al.}(2000)\citenamefont {Dvali},
  \citenamefont {Gabadadze},\ and\ \citenamefont {Porrati}}]{dvali2000}%
  \BibitemOpen
  \bibfield  {author} {\bibinfo {author} {\bibfnamefont {G.}~\bibnamefont
  {Dvali}}, \bibinfo {author} {\bibfnamefont {G.}~\bibnamefont {Gabadadze}}, \
  and\ \bibinfo {author} {\bibfnamefont {M.}~\bibnamefont {Porrati}},\
  }\href@noop {} {\bibfield  {journal} {\bibinfo  {journal} {Phys. Lett. B}\
  }\textbf {\bibinfo {volume} {485}},\ \bibinfo {pages} {208} (\bibinfo {year}
  {2000})}\BibitemShut {NoStop}%
\bibitem [{\citenamefont {Yin}\ \emph {et~al.}(2007)\citenamefont {Yin},
  \citenamefont {Wang}, \citenamefont {Abdalla},\ and\ \citenamefont
  {Lin}}]{yin2005}%
  \BibitemOpen
  \bibfield  {author} {\bibinfo {author} {\bibfnamefont {S.}~\bibnamefont
  {Yin}}, \bibinfo {author} {\bibfnamefont {B.}~\bibnamefont {Wang}}, \bibinfo
  {author} {\bibfnamefont {E.}~\bibnamefont {Abdalla}}, \ and\ \bibinfo
  {author} {\bibfnamefont {C.}~\bibnamefont {Lin}},\ }\href@noop {} {\bibfield
  {journal} {\bibinfo  {journal} {Phys.Rev. D}\ }\textbf {\bibinfo {volume}
  {76}},\ \bibinfo {pages} {124026} (\bibinfo {year} {2007})}\BibitemShut
  {NoStop}%
\bibitem [{\citenamefont {Peebles}\ and\ \citenamefont
  {Ratra}(1988)}]{peebles1988}%
  \BibitemOpen
  \bibfield  {author} {\bibinfo {author} {\bibfnamefont {P.~J.~E.}\
  \bibnamefont {Peebles}}\ and\ \bibinfo {author} {\bibfnamefont
  {B.}~\bibnamefont {Ratra}},\ }\href {\doibase 10.1086/185100} {\bibfield
  {journal} {\bibinfo  {journal} {Astrophys.J.}\ }\textbf {\bibinfo {volume}
  {325}},\ \bibinfo {pages} {L17} (\bibinfo {year} {1988})}\BibitemShut
  {NoStop}%
\bibitem [{\citenamefont {Ratra}\ and\ \citenamefont
  {Peebles}(1988)}]{ratra1988}%
  \BibitemOpen
  \bibfield  {author} {\bibinfo {author} {\bibfnamefont {B.}~\bibnamefont
  {Ratra}}\ and\ \bibinfo {author} {\bibfnamefont {P.~J.~E.}\ \bibnamefont
  {Peebles}},\ }\href {\doibase 10.1103/PhysRevD.37.3406} {\bibfield  {journal}
  {\bibinfo  {journal} {Phys.Rev.}\ }\textbf {\bibinfo {volume} {D37}},\
  \bibinfo {pages} {3406} (\bibinfo {year} {1988})}\BibitemShut {NoStop}%
\bibitem [{\citenamefont {Frieman}\ \emph {et~al.}(1992)\citenamefont
  {Frieman}, \citenamefont {Hill},\ and\ \citenamefont
  {Watkins}}]{Frieman1992}%
  \BibitemOpen
  \bibfield  {author} {\bibinfo {author} {\bibfnamefont {J.~A.}\ \bibnamefont
  {Frieman}}, \bibinfo {author} {\bibfnamefont {C.~T.}\ \bibnamefont {Hill}}, \
  and\ \bibinfo {author} {\bibfnamefont {R.}~\bibnamefont {Watkins}},\ }\href
  {\doibase 10.1103/PhysRevD.46.1226} {\bibfield  {journal} {\bibinfo
  {journal} {Phys.Rev.}\ }\textbf {\bibinfo {volume} {D46}},\ \bibinfo {pages}
  {1226} (\bibinfo {year} {1992})}\BibitemShut {NoStop}%
\bibitem [{\citenamefont {Frieman}\ \emph {et~al.}(1995)\citenamefont
  {Frieman}, \citenamefont {Hill}, \citenamefont {Stebbins},\ and\
  \citenamefont {Waga}}]{Frieman1995}%
  \BibitemOpen
  \bibfield  {author} {\bibinfo {author} {\bibfnamefont {J.~A.}\ \bibnamefont
  {Frieman}}, \bibinfo {author} {\bibfnamefont {C.~T.}\ \bibnamefont {Hill}},
  \bibinfo {author} {\bibfnamefont {A.}~\bibnamefont {Stebbins}}, \ and\
  \bibinfo {author} {\bibfnamefont {I.}~\bibnamefont {Waga}},\ }\href {\doibase
  10.1103/PhysRevLett.80.1582} {\bibfield  {journal} {\bibinfo  {journal}
  {Phys. Rev. Lett.}\ }\textbf {\bibinfo {volume} {75}},\ \bibinfo {pages}
  {2077} (\bibinfo {year} {1995})}\BibitemShut {NoStop}%
\bibitem [{\citenamefont {Caldwell}\ \emph {et~al.}(1998)\citenamefont
  {Caldwell}, \citenamefont {Dave},\ and\ \citenamefont
  {Steinhardt}}]{Caldwell:1997ii}%
  \BibitemOpen
  \bibfield  {author} {\bibinfo {author} {\bibfnamefont {R.~R.}\ \bibnamefont
  {Caldwell}}, \bibinfo {author} {\bibfnamefont {R.}~\bibnamefont {Dave}}, \
  and\ \bibinfo {author} {\bibfnamefont {P.~J.}\ \bibnamefont {Steinhardt}},\
  }\href {\doibase 10.1103/PhysRevLett.80.1582} {\bibfield  {journal} {\bibinfo
   {journal} {Phys. Rev. Lett.}\ }\textbf {\bibinfo {volume} {80}},\ \bibinfo
  {pages} {1582} (\bibinfo {year} {1998})}\BibitemShut {NoStop}%
\bibitem [{\citenamefont {Padmanabhan}(2002)}]{Padmanabhan:2002cp}%
  \BibitemOpen
  \bibfield  {author} {\bibinfo {author} {\bibfnamefont {T.}~\bibnamefont
  {Padmanabhan}},\ }\href {\doibase 10.1103/PhysRevD.66.021301} {\bibfield
  {journal} {\bibinfo  {journal} {Phys.Rev.}\ }\textbf {\bibinfo {volume}
  {D66}},\ \bibinfo {pages} {021301} (\bibinfo {year} {2002})},\ \Eprint
  {http://arxiv.org/abs/hep-th/0204150} {arXiv:hep-th/0204150 [hep-th]}
  \BibitemShut {NoStop}%
\bibitem [{\citenamefont {Bagla}\ \emph {et~al.}(2003)\citenamefont {Bagla},
  \citenamefont {Jassal},\ and\ \citenamefont {Padmanabhan}}]{Bagla:2002yn}%
  \BibitemOpen
  \bibfield  {author} {\bibinfo {author} {\bibfnamefont {J.~S.}\ \bibnamefont
  {Bagla}}, \bibinfo {author} {\bibfnamefont {H.~K.}\ \bibnamefont {Jassal}}, \
  and\ \bibinfo {author} {\bibfnamefont {T.}~\bibnamefont {Padmanabhan}},\
  }\href {\doibase 10.1103/PhysRevD.67.063504} {\bibfield  {journal} {\bibinfo
  {journal} {Phys.Rev.}\ }\textbf {\bibinfo {volume} {D67}},\ \bibinfo {pages}
  {063504} (\bibinfo {year} {2003})},\ \Eprint
  {http://arxiv.org/abs/astro-ph/0212198} {arXiv:astro-ph/0212198 [astro-ph]}
  \BibitemShut {NoStop}%
\bibitem [{\citenamefont {Armendariz-Picon}\ \emph {et~al.}(2000)\citenamefont
  {Armendariz-Picon}, \citenamefont {Mukhanov},\ and\ \citenamefont
  {Steinhardt}}]{ArmendarizPicon:2000dh}%
  \BibitemOpen
  \bibfield  {author} {\bibinfo {author} {\bibfnamefont {C.}~\bibnamefont
  {Armendariz-Picon}}, \bibinfo {author} {\bibfnamefont {V.~F.}\ \bibnamefont
  {Mukhanov}}, \ and\ \bibinfo {author} {\bibfnamefont {P.~J.}\ \bibnamefont
  {Steinhardt}},\ }\href {\doibase 10.1103/PhysRevLett.85.4438} {\bibfield
  {journal} {\bibinfo  {journal} {Phys. Rev. Lett.}\ }\textbf {\bibinfo
  {volume} {85}},\ \bibinfo {pages} {4438} (\bibinfo {year} {2000})},\ \Eprint
  {http://arxiv.org/abs/astro-ph/0004134} {arXiv:astro-ph/0004134 [astro-ph]}
  \BibitemShut {NoStop}%
\bibitem [{\citenamefont {Brax}\ and\ \citenamefont {Martin}(1999)}]{Brax1999}%
  \BibitemOpen
  \bibfield  {author} {\bibinfo {author} {\bibfnamefont {P.}~\bibnamefont
  {Brax}}\ and\ \bibinfo {author} {\bibfnamefont {J.}~\bibnamefont {Martin}},\
  }\href {\doibase 10.1016/S0370-2693(99)01209-5} {\bibfield  {journal}
  {\bibinfo  {journal} {Phys. Lett.}\ }\textbf {\bibinfo {volume} {B468}},\
  \bibinfo {pages} {40} (\bibinfo {year} {1999})},\ \Eprint
  {http://arxiv.org/abs/astro-ph/9905040} {arXiv:astro-ph/9905040 [astro-ph]}
  \BibitemShut {NoStop}%
\bibitem [{\citenamefont {Copeland}\ \emph {et~al.}(2000)\citenamefont
  {Copeland}, \citenamefont {Nunes},\ and\ \citenamefont
  {Rosati}}]{Copeland2000}%
  \BibitemOpen
  \bibfield  {author} {\bibinfo {author} {\bibfnamefont {E.~J.}\ \bibnamefont
  {Copeland}}, \bibinfo {author} {\bibfnamefont {N.~J.}\ \bibnamefont {Nunes}},
  \ and\ \bibinfo {author} {\bibfnamefont {F.}~\bibnamefont {Rosati}},\ }\href
  {\doibase 10.1103/PhysRevD.62.123503} {\bibfield  {journal} {\bibinfo
  {journal} {Phys. Rev.}\ }\textbf {\bibinfo {volume} {D62}},\ \bibinfo {pages}
  {123503} (\bibinfo {year} {2000})},\ \Eprint
  {http://arxiv.org/abs/hep-ph/0005222} {arXiv:hep-ph/0005222 [hep-ph]}
  \BibitemShut {NoStop}%
\bibitem [{\citenamefont {Landim}(2016{\natexlab{a}})}]{Landim:2015upa}%
  \BibitemOpen
  \bibfield  {author} {\bibinfo {author} {\bibfnamefont {R.~C.~G.}\
  \bibnamefont {Landim}},\ }\href {\doibase 10.1140/epjc/s10052-016-4287-2}
  {\bibfield  {journal} {\bibinfo  {journal} {Eur. Phys. J.}\ }\textbf
  {\bibinfo {volume} {C76}},\ \bibinfo {pages} {430} (\bibinfo {year}
  {2016}{\natexlab{a}})},\ \Eprint {http://arxiv.org/abs/1509.04980}
  {arXiv:1509.04980 [hep-th]} \BibitemShut {NoStop}%
\bibitem [{\citenamefont {Micheletti}\ \emph {et~al.}(2009)\citenamefont
  {Micheletti}, \citenamefont {Abdalla},\ and\ \citenamefont
  {Wang}}]{micheletti2009}%
  \BibitemOpen
  \bibfield  {author} {\bibinfo {author} {\bibfnamefont {S.}~\bibnamefont
  {Micheletti}}, \bibinfo {author} {\bibfnamefont {E.}~\bibnamefont {Abdalla}},
  \ and\ \bibinfo {author} {\bibfnamefont {B.}~\bibnamefont {Wang}},\ }\href
  {\doibase 10.1103/PhysRevD.79.123506} {\bibfield  {journal} {\bibinfo
  {journal} {Phys.Rev.}\ }\textbf {\bibinfo {volume} {D79}},\ \bibinfo {pages}
  {123506} (\bibinfo {year} {2009})},\ \Eprint {http://arxiv.org/abs/0902.0318}
  {arXiv:0902.0318 [gr-qc]} \BibitemShut {NoStop}%
\bibitem [{\citenamefont {Koivisto}\ and\ \citenamefont
  {Mota}(2008)}]{Koivisto:2008xf}%
  \BibitemOpen
  \bibfield  {author} {\bibinfo {author} {\bibfnamefont {T.}~\bibnamefont
  {Koivisto}}\ and\ \bibinfo {author} {\bibfnamefont {D.~F.}\ \bibnamefont
  {Mota}},\ }\href {\doibase 10.1088/1475-7516/2008/08/021} {\bibfield
  {journal} {\bibinfo  {journal} {JCAP}\ }\textbf {\bibinfo {volume} {0808}},\
  \bibinfo {pages} {021} (\bibinfo {year} {2008})},\ \Eprint
  {http://arxiv.org/abs/0805.4229} {arXiv:0805.4229 [astro-ph]} \BibitemShut
  {NoStop}%
\bibitem [{\citenamefont {Bamba}\ and\ \citenamefont
  {Odintsov}(2008)}]{Bamba:2008ja}%
  \BibitemOpen
  \bibfield  {author} {\bibinfo {author} {\bibfnamefont {K.}~\bibnamefont
  {Bamba}}\ and\ \bibinfo {author} {\bibfnamefont {S.~D.}\ \bibnamefont
  {Odintsov}},\ }\href {\doibase 10.1088/1475-7516/2008/04/024} {\bibfield
  {journal} {\bibinfo  {journal} {JCAP}\ }\textbf {\bibinfo {volume} {0804}},\
  \bibinfo {pages} {024} (\bibinfo {year} {2008})},\ \Eprint
  {http://arxiv.org/abs/0801.0954} {arXiv:0801.0954 [astro-ph]} \BibitemShut
  {NoStop}%
\bibitem [{\citenamefont {Emelyanov}\ and\ \citenamefont
  {Klinkhamer}(2012{\natexlab{a}})}]{Emelyanov:2011ze}%
  \BibitemOpen
  \bibfield  {author} {\bibinfo {author} {\bibfnamefont {V.}~\bibnamefont
  {Emelyanov}}\ and\ \bibinfo {author} {\bibfnamefont {F.~R.}\ \bibnamefont
  {Klinkhamer}},\ }\href {\doibase 10.1103/PhysRevD.85.103508} {\bibfield
  {journal} {\bibinfo  {journal} {Phys. Rev.}\ }\textbf {\bibinfo {volume}
  {D85}},\ \bibinfo {pages} {103508} (\bibinfo {year} {2012}{\natexlab{a}})},\
  \Eprint {http://arxiv.org/abs/1109.4915} {arXiv:1109.4915 [hep-th]}
  \BibitemShut {NoStop}%
\bibitem [{\citenamefont {Emelyanov}\ and\ \citenamefont
  {Klinkhamer}(2012{\natexlab{b}})}]{Emelyanov:2011wn}%
  \BibitemOpen
  \bibfield  {author} {\bibinfo {author} {\bibfnamefont {V.}~\bibnamefont
  {Emelyanov}}\ and\ \bibinfo {author} {\bibfnamefont {F.~R.}\ \bibnamefont
  {Klinkhamer}},\ }\href {\doibase 10.1103/PhysRevD.85.063522} {\bibfield
  {journal} {\bibinfo  {journal} {Phys. Rev.}\ }\textbf {\bibinfo {volume}
  {D85}},\ \bibinfo {pages} {063522} (\bibinfo {year} {2012}{\natexlab{b}})},\
  \Eprint {http://arxiv.org/abs/1107.0961} {arXiv:1107.0961 [hep-th]}
  \BibitemShut {NoStop}%
\bibitem [{\citenamefont {Emelyanov}\ and\ \citenamefont
  {Klinkhamer}(2012{\natexlab{c}})}]{Emelyanov:2011kn}%
  \BibitemOpen
  \bibfield  {author} {\bibinfo {author} {\bibfnamefont {V.}~\bibnamefont
  {Emelyanov}}\ and\ \bibinfo {author} {\bibfnamefont {F.~R.}\ \bibnamefont
  {Klinkhamer}},\ }\href {\doibase 10.1142/S0218271812500253} {\bibfield
  {journal} {\bibinfo  {journal} {Int. J. Mod. Phys.}\ }\textbf {\bibinfo
  {volume} {D21}},\ \bibinfo {pages} {1250025} (\bibinfo {year}
  {2012}{\natexlab{c}})},\ \Eprint {http://arxiv.org/abs/1108.1995}
  {arXiv:1108.1995 [gr-qc]} \BibitemShut {NoStop}%
\bibitem [{\citenamefont {Kouwn}\ \emph {et~al.}(2016)\citenamefont {Kouwn},
  \citenamefont {Oh},\ and\ \citenamefont {Park}}]{Kouwn:2015cdw}%
  \BibitemOpen
  \bibfield  {author} {\bibinfo {author} {\bibfnamefont {S.}~\bibnamefont
  {Kouwn}}, \bibinfo {author} {\bibfnamefont {P.}~\bibnamefont {Oh}}, \ and\
  \bibinfo {author} {\bibfnamefont {C.-G.}\ \bibnamefont {Park}},\ }\href
  {\doibase 10.1103/PhysRevD.93.083012} {\bibfield  {journal} {\bibinfo
  {journal} {Phys. Rev.}\ }\textbf {\bibinfo {volume} {D93}},\ \bibinfo {pages}
  {083012} (\bibinfo {year} {2016})},\ \Eprint
  {http://arxiv.org/abs/1512.00541} {arXiv:1512.00541 [astro-ph.CO]}
  \BibitemShut {NoStop}%
\bibitem [{\citenamefont {Landim}(2016{\natexlab{b}})}]{Landim:2016dxh}%
  \BibitemOpen
  \bibfield  {author} {\bibinfo {author} {\bibfnamefont {R.~C.~G.}\
  \bibnamefont {Landim}},\ }\href {\doibase 10.1140/epjc/s10052-016-4328-x}
  {\bibfield  {journal} {\bibinfo  {journal} {Eur. Phys. J.}\ }\textbf
  {\bibinfo {volume} {C76}},\ \bibinfo {pages} {480} (\bibinfo {year}
  {2016}{\natexlab{b}})},\ \Eprint {http://arxiv.org/abs/1605.03550}
  {arXiv:1605.03550 [gr-qc]} \BibitemShut {NoStop}%
\bibitem [{\citenamefont {Costa}\ \emph
  {et~al.}(2014{\natexlab{a}})\citenamefont {Costa}, \citenamefont {Olivari},\
  and\ \citenamefont {Abdalla}}]{costa2014}%
  \BibitemOpen
  \bibfield  {author} {\bibinfo {author} {\bibfnamefont {A.~A.}\ \bibnamefont
  {Costa}}, \bibinfo {author} {\bibfnamefont {L.~C.}\ \bibnamefont {Olivari}},
  \ and\ \bibinfo {author} {\bibfnamefont {E.}~\bibnamefont {Abdalla}},\
  }\href@noop {} {\bibfield  {journal} {\bibinfo  {journal} {Phys. Rev.}\
  }\textbf {\bibinfo {volume} {D92}},\ \bibinfo {pages} {103501} (\bibinfo
  {year} {2014}{\natexlab{a}})},\ \Eprint {http://arxiv.org/abs/1411.3660}
  {arXiv:1411.3660 [hep-th]} \BibitemShut {NoStop}%
\bibitem [{\citenamefont {Hsu}(2004)}]{Hsu:2004ri}%
  \BibitemOpen
  \bibfield  {author} {\bibinfo {author} {\bibfnamefont {S.~D.~H.}\
  \bibnamefont {Hsu}},\ }\href {\doibase 10.1016/j.physletb.2004.05.020}
  {\bibfield  {journal} {\bibinfo  {journal} {Phys. Lett.}\ }\textbf {\bibinfo
  {volume} {B594}},\ \bibinfo {pages} {13} (\bibinfo {year} {2004})},\ \Eprint
  {http://arxiv.org/abs/hep-th/0403052} {arXiv:hep-th/0403052 [hep-th]}
  \BibitemShut {NoStop}%
\bibitem [{\citenamefont {Li}(2004)}]{Li:2004rb}%
  \BibitemOpen
  \bibfield  {author} {\bibinfo {author} {\bibfnamefont {M.}~\bibnamefont
  {Li}},\ }\href {\doibase 10.1016/j.physletb.2004.10.014} {\bibfield
  {journal} {\bibinfo  {journal} {Phys. Lett.}\ }\textbf {\bibinfo {volume}
  {B603}},\ \bibinfo {pages} {1} (\bibinfo {year} {2004})},\ \Eprint
  {http://arxiv.org/abs/hep-th/0403127} {arXiv:hep-th/0403127 [hep-th]}
  \BibitemShut {NoStop}%
\bibitem [{\citenamefont {Pavon}\ and\ \citenamefont
  {Zimdahl}(2005)}]{Pavon:2005yx}%
  \BibitemOpen
  \bibfield  {author} {\bibinfo {author} {\bibfnamefont {D.}~\bibnamefont
  {Pavon}}\ and\ \bibinfo {author} {\bibfnamefont {W.}~\bibnamefont
  {Zimdahl}},\ }\href {\doibase 10.1016/j.physletb.2005.08.134} {\bibfield
  {journal} {\bibinfo  {journal} {Phys. Lett.}\ }\textbf {\bibinfo {volume}
  {B628}},\ \bibinfo {pages} {206} (\bibinfo {year} {2005})},\ \Eprint
  {http://arxiv.org/abs/gr-qc/0505020} {arXiv:gr-qc/0505020 [gr-qc]}
  \BibitemShut {NoStop}%
\bibitem [{\citenamefont {Wang}\ \emph {et~al.}(2005)\citenamefont {Wang},
  \citenamefont {Gong},\ and\ \citenamefont {Abdalla}}]{Wang:2005jx}%
  \BibitemOpen
  \bibfield  {author} {\bibinfo {author} {\bibfnamefont {B.}~\bibnamefont
  {Wang}}, \bibinfo {author} {\bibfnamefont {Y.-G.}\ \bibnamefont {Gong}}, \
  and\ \bibinfo {author} {\bibfnamefont {E.}~\bibnamefont {Abdalla}},\ }\href
  {\doibase 10.1016/j.physletb.2005.08.008} {\bibfield  {journal} {\bibinfo
  {journal} {Phys. Lett.}\ }\textbf {\bibinfo {volume} {B624}},\ \bibinfo
  {pages} {141} (\bibinfo {year} {2005})},\ \Eprint
  {http://arxiv.org/abs/hep-th/0506069} {arXiv:hep-th/0506069 [hep-th]}
  \BibitemShut {NoStop}%
\bibitem [{\citenamefont {Wang}\ \emph
  {et~al.}(2006{\natexlab{a}})\citenamefont {Wang}, \citenamefont {Gong},\ and\
  \citenamefont {Abdalla}}]{Wang:2005pk}%
  \BibitemOpen
  \bibfield  {author} {\bibinfo {author} {\bibfnamefont {B.}~\bibnamefont
  {Wang}}, \bibinfo {author} {\bibfnamefont {Y.}~\bibnamefont {Gong}}, \ and\
  \bibinfo {author} {\bibfnamefont {E.}~\bibnamefont {Abdalla}},\ }\href
  {\doibase 10.1103/PhysRevD.74.083520} {\bibfield  {journal} {\bibinfo
  {journal} {Phys. Rev.}\ }\textbf {\bibinfo {volume} {D74}},\ \bibinfo {pages}
  {083520} (\bibinfo {year} {2006}{\natexlab{a}})},\ \Eprint
  {http://arxiv.org/abs/gr-qc/0511051} {arXiv:gr-qc/0511051 [gr-qc]}
  \BibitemShut {NoStop}%
\bibitem [{\citenamefont {Wang}\ \emph
  {et~al.}(2006{\natexlab{b}})\citenamefont {Wang}, \citenamefont {Lin},\ and\
  \citenamefont {Abdalla}}]{Wang:2005ph}%
  \BibitemOpen
  \bibfield  {author} {\bibinfo {author} {\bibfnamefont {B.}~\bibnamefont
  {Wang}}, \bibinfo {author} {\bibfnamefont {C.-Y.}\ \bibnamefont {Lin}}, \
  and\ \bibinfo {author} {\bibfnamefont {E.}~\bibnamefont {Abdalla}},\ }\href
  {\doibase 10.1016/j.physletb.2006.04.009} {\bibfield  {journal} {\bibinfo
  {journal} {Phys. Lett.}\ }\textbf {\bibinfo {volume} {B637}},\ \bibinfo
  {pages} {357} (\bibinfo {year} {2006}{\natexlab{b}})},\ \Eprint
  {http://arxiv.org/abs/hep-th/0509107} {arXiv:hep-th/0509107 [hep-th]}
  \BibitemShut {NoStop}%
\bibitem [{\citenamefont {Wang}\ \emph {et~al.}(2008)\citenamefont {Wang},
  \citenamefont {Lin}, \citenamefont {Pavon},\ and\ \citenamefont
  {Abdalla}}]{Wang:2007ak}%
  \BibitemOpen
  \bibfield  {author} {\bibinfo {author} {\bibfnamefont {B.}~\bibnamefont
  {Wang}}, \bibinfo {author} {\bibfnamefont {C.-Y.}\ \bibnamefont {Lin}},
  \bibinfo {author} {\bibfnamefont {D.}~\bibnamefont {Pavon}}, \ and\ \bibinfo
  {author} {\bibfnamefont {E.}~\bibnamefont {Abdalla}},\ }\href {\doibase
  10.1016/j.physletb.2008.01.074} {\bibfield  {journal} {\bibinfo  {journal}
  {Phys. Lett.}\ }\textbf {\bibinfo {volume} {B662}},\ \bibinfo {pages} {1}
  (\bibinfo {year} {2008})},\ \Eprint {http://arxiv.org/abs/0711.2214}
  {arXiv:0711.2214 [hep-th]} \BibitemShut {NoStop}%
\bibitem [{\citenamefont {Landim}(2016{\natexlab{c}})}]{Landim:2015hqa}%
  \BibitemOpen
  \bibfield  {author} {\bibinfo {author} {\bibfnamefont {R.~C.~G.}\
  \bibnamefont {Landim}},\ }\href {\doibase 10.1142/S0218271816500504}
  {\bibfield  {journal} {\bibinfo  {journal} {Int. J. Mod. Phys.}\ }\textbf
  {\bibinfo {volume} {D25}},\ \bibinfo {pages} {1650050} (\bibinfo {year}
  {2016}{\natexlab{c}})},\ \Eprint {http://arxiv.org/abs/1508.07248}
  {arXiv:1508.07248 [hep-th]} \BibitemShut {NoStop}%
\bibitem [{\citenamefont {Dymnikova}\ and\ \citenamefont
  {Khlopov}(2000)}]{Dymnikova:2001ga}%
  \BibitemOpen
  \bibfield  {author} {\bibinfo {author} {\bibfnamefont {I.}~\bibnamefont
  {Dymnikova}}\ and\ \bibinfo {author} {\bibfnamefont {M.}~\bibnamefont
  {Khlopov}},\ }\href {\doibase 10.1142/S0217732300002966} {\bibfield
  {journal} {\bibinfo  {journal} {Mod. Phys. Lett.}\ }\textbf {\bibinfo
  {volume} {A15}},\ \bibinfo {pages} {2305} (\bibinfo {year} {2000})},\ \Eprint
  {http://arxiv.org/abs/astro-ph/0102094} {arXiv:astro-ph/0102094 [astro-ph]}
  \BibitemShut {NoStop}%
\bibitem [{\citenamefont {Dymnikova}\ and\ \citenamefont
  {Khlopov}(2001)}]{Dymnikova:2001jy}%
  \BibitemOpen
  \bibfield  {author} {\bibinfo {author} {\bibfnamefont {I.}~\bibnamefont
  {Dymnikova}}\ and\ \bibinfo {author} {\bibfnamefont {M.}~\bibnamefont
  {Khlopov}},\ }\href {\doibase 10.1007/s100520100625} {\bibfield  {journal}
  {\bibinfo  {journal} {Eur. Phys. J.}\ }\textbf {\bibinfo {volume} {C20}},\
  \bibinfo {pages} {139} (\bibinfo {year} {2001})}\BibitemShut {NoStop}%
\bibitem [{\citenamefont {Mukhopadhyay}\ \emph {et~al.}(2011)\citenamefont
  {Mukhopadhyay}, \citenamefont {Ghosh}, \citenamefont {Khlopov},\ and\
  \citenamefont {Ray}}]{Mukhopadhyay:2007ed}%
  \BibitemOpen
  \bibfield  {author} {\bibinfo {author} {\bibfnamefont {U.}~\bibnamefont
  {Mukhopadhyay}}, \bibinfo {author} {\bibfnamefont {P.~P.}\ \bibnamefont
  {Ghosh}}, \bibinfo {author} {\bibfnamefont {M.}~\bibnamefont {Khlopov}}, \
  and\ \bibinfo {author} {\bibfnamefont {S.}~\bibnamefont {Ray}},\ }\href
  {\doibase 10.1007/s10773-010-0639-0} {\bibfield  {journal} {\bibinfo
  {journal} {Int. J. Theor. Phys.}\ }\textbf {\bibinfo {volume} {50}},\
  \bibinfo {pages} {939} (\bibinfo {year} {2011})},\ \Eprint
  {http://arxiv.org/abs/0711.0686} {arXiv:0711.0686 [gr-qc]} \BibitemShut
  {NoStop}%
\bibitem [{\citenamefont {Wetterich}(1995)}]{Wetterich:1994bg}%
  \BibitemOpen
  \bibfield  {author} {\bibinfo {author} {\bibfnamefont {C.}~\bibnamefont
  {Wetterich}},\ }\href@noop {} {\bibfield  {journal} {\bibinfo  {journal}
  {Astron.Astrophys.}\ }\textbf {\bibinfo {volume} {301}},\ \bibinfo {pages}
  {321} (\bibinfo {year} {1995})},\ \Eprint
  {http://arxiv.org/abs/hep-th/9408025} {arXiv:hep-th/9408025 [hep-th]}
  \BibitemShut {NoStop}%
\bibitem [{\citenamefont {Amendola}(2000)}]{Amendola:1999er}%
  \BibitemOpen
  \bibfield  {author} {\bibinfo {author} {\bibfnamefont {L.}~\bibnamefont
  {Amendola}},\ }\href {\doibase 10.1103/PhysRevD.62.043511} {\bibfield
  {journal} {\bibinfo  {journal} {Phys.Rev.}\ }\textbf {\bibinfo {volume}
  {D62}},\ \bibinfo {pages} {043511} (\bibinfo {year} {2000})},\ \Eprint
  {http://arxiv.org/abs/astro-ph/9908023} {arXiv:astro-ph/9908023 [astro-ph]}
  \BibitemShut {NoStop}%
\bibitem [{\citenamefont {Wang}\ \emph {et~al.}(2016)\citenamefont {Wang},
  \citenamefont {Abdalla}, \citenamefont {Atrio-Barandela},\ and\ \citenamefont
  {Pavon}}]{Wang:2016lxa}%
  \BibitemOpen
  \bibfield  {author} {\bibinfo {author} {\bibfnamefont {B.}~\bibnamefont
  {Wang}}, \bibinfo {author} {\bibfnamefont {E.}~\bibnamefont {Abdalla}},
  \bibinfo {author} {\bibfnamefont {F.}~\bibnamefont {Atrio-Barandela}}, \ and\
  \bibinfo {author} {\bibfnamefont {D.}~\bibnamefont {Pavon}},\ }\href
  {\doibase 10.1088/0034-4885/79/9/096901} {\bibfield  {journal} {\bibinfo
  {journal} {Rep. Prog. Phys.}\ }\textbf {\bibinfo {volume} {79}},\ \bibinfo
  {pages} {096901} (\bibinfo {year} {2016})},\ \Eprint
  {http://arxiv.org/abs/1603.08299} {arXiv:1603.08299 [astro-ph.CO]}
  \BibitemShut {NoStop}%
\bibitem [{\citenamefont {Zimdahl}\ and\ \citenamefont
  {Pavon}(2001)}]{Zimdahl:2001ar}%
  \BibitemOpen
  \bibfield  {author} {\bibinfo {author} {\bibfnamefont {W.}~\bibnamefont
  {Zimdahl}}\ and\ \bibinfo {author} {\bibfnamefont {D.}~\bibnamefont
  {Pavon}},\ }\href {\doibase 10.1016/S0370-2693(01)01174-1} {\bibfield
  {journal} {\bibinfo  {journal} {Phys.Lett.}\ }\textbf {\bibinfo {volume}
  {B521}},\ \bibinfo {pages} {133} (\bibinfo {year} {2001})},\ \Eprint
  {http://arxiv.org/abs/astro-ph/0105479} {arXiv:astro-ph/0105479 [astro-ph]}
  \BibitemShut {NoStop}%
\bibitem [{\citenamefont {Chimento}\ \emph {et~al.}(2003)\citenamefont
  {Chimento}, \citenamefont {Jakubi}, \citenamefont {Pavon},\ and\
  \citenamefont {Zimdahl}}]{Chimento:2003iea}%
  \BibitemOpen
  \bibfield  {author} {\bibinfo {author} {\bibfnamefont {L.~P.}\ \bibnamefont
  {Chimento}}, \bibinfo {author} {\bibfnamefont {A.~S.}\ \bibnamefont
  {Jakubi}}, \bibinfo {author} {\bibfnamefont {D.}~\bibnamefont {Pavon}}, \
  and\ \bibinfo {author} {\bibfnamefont {W.}~\bibnamefont {Zimdahl}},\ }\href
  {\doibase 10.1103/PhysRevD.67.083513} {\bibfield  {journal} {\bibinfo
  {journal} {Phys.Rev.}\ }\textbf {\bibinfo {volume} {D67}},\ \bibinfo {pages}
  {083513} (\bibinfo {year} {2003})},\ \Eprint
  {http://arxiv.org/abs/astro-ph/0303145} {arXiv:astro-ph/0303145 [astro-ph]}
  \BibitemShut {NoStop}%
\bibitem [{\citenamefont {Guo}\ and\ \citenamefont {Zhang}(2005)}]{Guo:2004vg}%
  \BibitemOpen
  \bibfield  {author} {\bibinfo {author} {\bibfnamefont {Z.-K.}\ \bibnamefont
  {Guo}}\ and\ \bibinfo {author} {\bibfnamefont {Y.-Z.}\ \bibnamefont
  {Zhang}},\ }\href {\doibase 10.1103/PhysRevD.71.023501} {\bibfield  {journal}
  {\bibinfo  {journal} {Phys. Rev. D.}\ }\textbf {\bibinfo {volume} {71}},\
  \bibinfo {pages} {023501} (\bibinfo {year} {2005})},\ \Eprint
  {http://arxiv.org/abs/astro-ph/0411524} {arXiv:astro-ph/0411524 [astro-ph]}
  \BibitemShut {NoStop}%
\bibitem [{\citenamefont {Cai}\ and\ \citenamefont {Wang}(2005)}]{Cai:2004dk}%
  \BibitemOpen
  \bibfield  {author} {\bibinfo {author} {\bibfnamefont {R.-G.}\ \bibnamefont
  {Cai}}\ and\ \bibinfo {author} {\bibfnamefont {A.}~\bibnamefont {Wang}},\
  }\href {\doibase 10.1088/1475-7516/2005/03/002} {\bibfield  {journal}
  {\bibinfo  {journal} {JCAP}\ }\textbf {\bibinfo {volume} {0503}},\ \bibinfo
  {pages} {002} (\bibinfo {year} {2005})},\ \Eprint
  {http://arxiv.org/abs/hep-th/0411025} {arXiv:hep-th/0411025 [hep-th]}
  \BibitemShut {NoStop}%
\bibitem [{\citenamefont {Guo}\ \emph {et~al.}(2005)\citenamefont {Guo},
  \citenamefont {Cai},\ and\ \citenamefont {Zhang}}]{Guo:2004xx}%
  \BibitemOpen
  \bibfield  {author} {\bibinfo {author} {\bibfnamefont {Z.-K.}\ \bibnamefont
  {Guo}}, \bibinfo {author} {\bibfnamefont {R.-G.}\ \bibnamefont {Cai}}, \ and\
  \bibinfo {author} {\bibfnamefont {Y.-Z.}\ \bibnamefont {Zhang}},\ }\href
  {\doibase 10.1088/1475-7516/2005/05/002} {\bibfield  {journal} {\bibinfo
  {journal} {JCAP}\ }\textbf {\bibinfo {volume} {0505}},\ \bibinfo {pages}
  {002} (\bibinfo {year} {2005})},\ \Eprint
  {http://arxiv.org/abs/astro-ph/0412624} {arXiv:astro-ph/0412624 [astro-ph]}
  \BibitemShut {NoStop}%
\bibitem [{\citenamefont {Bi}\ \emph {et~al.}(2005)\citenamefont {Bi},
  \citenamefont {Feng}, \citenamefont {Li},\ and\ \citenamefont
  {Zhang}}]{Bi:2004ns}%
  \BibitemOpen
  \bibfield  {author} {\bibinfo {author} {\bibfnamefont {X.-J.}\ \bibnamefont
  {Bi}}, \bibinfo {author} {\bibfnamefont {B.}~\bibnamefont {Feng}}, \bibinfo
  {author} {\bibfnamefont {H.}~\bibnamefont {Li}}, \ and\ \bibinfo {author}
  {\bibfnamefont {X.}~\bibnamefont {Zhang}},\ }\href {\doibase
  10.1103/PhysRevD.72.123523} {\bibfield  {journal} {\bibinfo  {journal} {Phys.
  Rev. D.}\ }\textbf {\bibinfo {volume} {72}},\ \bibinfo {pages} {123523}
  (\bibinfo {year} {2005})},\ \Eprint {http://arxiv.org/abs/hep-ph/0412002}
  {arXiv:hep-ph/0412002 [hep-ph]} \BibitemShut {NoStop}%
\bibitem [{\citenamefont {Gumjudpai}\ \emph {et~al.}(2005)\citenamefont
  {Gumjudpai}, \citenamefont {Naskar}, \citenamefont {Sami},\ and\
  \citenamefont {Tsujikawa}}]{Gumjudpai:2005ry}%
  \BibitemOpen
  \bibfield  {author} {\bibinfo {author} {\bibfnamefont {B.}~\bibnamefont
  {Gumjudpai}}, \bibinfo {author} {\bibfnamefont {T.}~\bibnamefont {Naskar}},
  \bibinfo {author} {\bibfnamefont {M.}~\bibnamefont {Sami}}, \ and\ \bibinfo
  {author} {\bibfnamefont {S.}~\bibnamefont {Tsujikawa}},\ }\href {\doibase
  10.1088/1475-7516/2005/06/007} {\bibfield  {journal} {\bibinfo  {journal}
  {JCAP}\ }\textbf {\bibinfo {volume} {0506}},\ \bibinfo {pages} {007}
  (\bibinfo {year} {2005})},\ \Eprint {http://arxiv.org/abs/hep-th/0502191}
  {arXiv:hep-th/0502191 [hep-th]} \BibitemShut {NoStop}%
\bibitem [{\citenamefont {Costa}\ \emph
  {et~al.}(2014{\natexlab{b}})\citenamefont {Costa}, \citenamefont {Xu},
  \citenamefont {Wang}, \citenamefont {Ferreira},\ and\ \citenamefont
  {Abdalla}}]{Costa:2013sva}%
  \BibitemOpen
  \bibfield  {author} {\bibinfo {author} {\bibfnamefont {A.~A.}\ \bibnamefont
  {Costa}}, \bibinfo {author} {\bibfnamefont {X.-D.}\ \bibnamefont {Xu}},
  \bibinfo {author} {\bibfnamefont {B.}~\bibnamefont {Wang}}, \bibinfo {author}
  {\bibfnamefont {E.~G.~M.}\ \bibnamefont {Ferreira}}, \ and\ \bibinfo {author}
  {\bibfnamefont {E.}~\bibnamefont {Abdalla}},\ }\href {\doibase
  10.1103/PhysRevD.89.103531} {\bibfield  {journal} {\bibinfo  {journal} {Phys.
  Rev.}\ }\textbf {\bibinfo {volume} {D89}},\ \bibinfo {pages} {103531}
  (\bibinfo {year} {2014}{\natexlab{b}})},\ \Eprint
  {http://arxiv.org/abs/1311.7380} {arXiv:1311.7380 [astro-ph.CO]} \BibitemShut
  {NoStop}%
\bibitem [{\citenamefont {Abdalla}\ \emph {et~al.}(2014)\citenamefont
  {Abdalla}, \citenamefont {Ferreira}, \citenamefont {Quintin},\ and\
  \citenamefont {Wang}}]{Abdalla:2014cla}%
  \BibitemOpen
  \bibfield  {author} {\bibinfo {author} {\bibfnamefont {E.}~\bibnamefont
  {Abdalla}}, \bibinfo {author} {\bibfnamefont {E.~G.~M.}\ \bibnamefont
  {Ferreira}}, \bibinfo {author} {\bibfnamefont {J.}~\bibnamefont {Quintin}}, \
  and\ \bibinfo {author} {\bibfnamefont {B.}~\bibnamefont {Wang}},\ }\href@noop
  {} {\  (\bibinfo {year} {2014})},\ \Eprint {http://arxiv.org/abs/1412.2777}
  {arXiv:1412.2777 [astro-ph.CO]} \BibitemShut {NoStop}%
\bibitem [{\citenamefont {Costa}\ \emph {et~al.}(2015)\citenamefont {Costa},
  \citenamefont {Olivari},\ and\ \citenamefont {Abdalla}}]{Costa:2014pba}%
  \BibitemOpen
  \bibfield  {author} {\bibinfo {author} {\bibfnamefont {A.~A.}\ \bibnamefont
  {Costa}}, \bibinfo {author} {\bibfnamefont {L.~C.}\ \bibnamefont {Olivari}},
  \ and\ \bibinfo {author} {\bibfnamefont {E.}~\bibnamefont {Abdalla}},\ }\href
  {\doibase 10.1103/PhysRevD.92.103501} {\bibfield  {journal} {\bibinfo
  {journal} {Phys. Rev.}\ }\textbf {\bibinfo {volume} {D92}},\ \bibinfo {pages}
  {103501} (\bibinfo {year} {2015})},\ \Eprint {http://arxiv.org/abs/1411.3660}
  {arXiv:1411.3660 [astro-ph.CO]} \BibitemShut {NoStop}%
\bibitem [{\citenamefont {Costa}\ \emph {et~al.}(2016)\citenamefont {Costa},
  \citenamefont {Xu}, \citenamefont {Wang},\ and\ \citenamefont
  {Abdalla}}]{Costa:2016tpb}%
  \BibitemOpen
  \bibfield  {author} {\bibinfo {author} {\bibfnamefont {A.~A.}\ \bibnamefont
  {Costa}}, \bibinfo {author} {\bibfnamefont {X.-D.}\ \bibnamefont {Xu}},
  \bibinfo {author} {\bibfnamefont {B.}~\bibnamefont {Wang}}, \ and\ \bibinfo
  {author} {\bibfnamefont {E.}~\bibnamefont {Abdalla}},\ }\href@noop {} {\
  (\bibinfo {year} {2016})},\ \Eprint {http://arxiv.org/abs/1605.04138}
  {arXiv:1605.04138 [astro-ph.CO]} \BibitemShut {NoStop}%
\bibitem [{\citenamefont {Marcondes}\ \emph {et~al.}(2016)\citenamefont
  {Marcondes}, \citenamefont {Landim}, \citenamefont {Costa}, \citenamefont
  {Wang},\ and\ \citenamefont {Abdalla}}]{Marcondes:2016reb}%
  \BibitemOpen
  \bibfield  {author} {\bibinfo {author} {\bibfnamefont {R.~J.~F.}\
  \bibnamefont {Marcondes}}, \bibinfo {author} {\bibfnamefont {R.~C.~G.}\
  \bibnamefont {Landim}}, \bibinfo {author} {\bibfnamefont {A.~A.}\
  \bibnamefont {Costa}}, \bibinfo {author} {\bibfnamefont {B.}~\bibnamefont
  {Wang}}, \ and\ \bibinfo {author} {\bibfnamefont {E.}~\bibnamefont
  {Abdalla}},\ }\href@noop {} {\  (\bibinfo {year} {2016})},\ \Eprint
  {http://arxiv.org/abs/1605.05264} {arXiv:1605.05264 [astro-ph.CO]}
  \BibitemShut {NoStop}%
\bibitem [{\citenamefont {Farrar}\ and\ \citenamefont
  {Peebles}(2004)}]{Farrar:2003uw}%
  \BibitemOpen
  \bibfield  {author} {\bibinfo {author} {\bibfnamefont {G.~R.}\ \bibnamefont
  {Farrar}}\ and\ \bibinfo {author} {\bibfnamefont {P.~J.~E.}\ \bibnamefont
  {Peebles}},\ }\href {\doibase 10.1086/381728} {\bibfield  {journal} {\bibinfo
   {journal} {Astrophys. J.}\ }\textbf {\bibinfo {volume} {604}},\ \bibinfo
  {pages} {1} (\bibinfo {year} {2004})},\ \Eprint
  {http://arxiv.org/abs/astro-ph/0307316} {arXiv:astro-ph/0307316 [astro-ph]}
  \BibitemShut {NoStop}%
\bibitem [{\citenamefont {Abdalla}\ \emph {et~al.}(2013)\citenamefont
  {Abdalla}, \citenamefont {Graef},\ and\ \citenamefont
  {Wang}}]{Abdalla:2012ug}%
  \BibitemOpen
  \bibfield  {author} {\bibinfo {author} {\bibfnamefont {E.}~\bibnamefont
  {Abdalla}}, \bibinfo {author} {\bibfnamefont {L.~L.}\ \bibnamefont {Graef}},
  \ and\ \bibinfo {author} {\bibfnamefont {B.}~\bibnamefont {Wang}},\ }\href
  {\doibase 10.1016/j.physletb.2013.08.011} {\bibfield  {journal} {\bibinfo
  {journal} {Phys. Lett.}\ }\textbf {\bibinfo {volume} {B726}},\ \bibinfo
  {pages} {786} (\bibinfo {year} {2013})},\ \Eprint
  {http://arxiv.org/abs/1202.0499} {arXiv:1202.0499 [gr-qc]} \BibitemShut
  {NoStop}%
\bibitem [{\citenamefont {D'Amico}\ \emph {et~al.}(2016)\citenamefont
  {D'Amico}, \citenamefont {Hamill},\ and\ \citenamefont
  {Kaloper}}]{D'Amico:2016kqm}%
  \BibitemOpen
  \bibfield  {author} {\bibinfo {author} {\bibfnamefont {G.}~\bibnamefont
  {D'Amico}}, \bibinfo {author} {\bibfnamefont {T.}~\bibnamefont {Hamill}}, \
  and\ \bibinfo {author} {\bibfnamefont {N.}~\bibnamefont {Kaloper}},\
  }\href@noop {} {\  (\bibinfo {year} {2016})},\ \Eprint
  {http://arxiv.org/abs/1605.00996} {arXiv:1605.00996 [hep-th]} \BibitemShut
  {NoStop}%
\bibitem [{\citenamefont {Stojkovic}\ \emph {et~al.}(2008)\citenamefont
  {Stojkovic}, \citenamefont {Starkman},\ and\ \citenamefont
  {Matsuo}}]{Stojkovic:2007dw}%
  \BibitemOpen
  \bibfield  {author} {\bibinfo {author} {\bibfnamefont {D.}~\bibnamefont
  {Stojkovic}}, \bibinfo {author} {\bibfnamefont {G.~D.}\ \bibnamefont
  {Starkman}}, \ and\ \bibinfo {author} {\bibfnamefont {R.}~\bibnamefont
  {Matsuo}},\ }\href {\doibase 10.1103/PhysRevD.77.063006} {\bibfield
  {journal} {\bibinfo  {journal} {Phys. Rev.}\ }\textbf {\bibinfo {volume}
  {D77}},\ \bibinfo {pages} {063006} (\bibinfo {year} {2008})},\ \Eprint
  {http://arxiv.org/abs/hep-ph/0703246} {arXiv:hep-ph/0703246 [hep-ph]}
  \BibitemShut {NoStop}%
\bibitem [{\citenamefont {Greenwood}\ \emph {et~al.}(2009)\citenamefont
  {Greenwood}, \citenamefont {Halstead}, \citenamefont {Poltis},\ and\
  \citenamefont {Stojkovic}}]{Greenwood:2008qp}%
  \BibitemOpen
  \bibfield  {author} {\bibinfo {author} {\bibfnamefont {E.}~\bibnamefont
  {Greenwood}}, \bibinfo {author} {\bibfnamefont {E.}~\bibnamefont {Halstead}},
  \bibinfo {author} {\bibfnamefont {R.}~\bibnamefont {Poltis}}, \ and\ \bibinfo
  {author} {\bibfnamefont {D.}~\bibnamefont {Stojkovic}},\ }\href {\doibase
  10.1103/PhysRevD.79.103003} {\bibfield  {journal} {\bibinfo  {journal} {Phys.
  Rev.}\ }\textbf {\bibinfo {volume} {D79}},\ \bibinfo {pages} {103003}
  (\bibinfo {year} {2009})},\ \Eprint {http://arxiv.org/abs/0810.5343}
  {arXiv:0810.5343 [hep-ph]} \BibitemShut {NoStop}%
\bibitem [{\citenamefont {Shafieloo}\ \emph {et~al.}(2016)\citenamefont
  {Shafieloo}, \citenamefont {Hazra}, \citenamefont {Sahni},\ and\
  \citenamefont {Starobinsky}}]{Shafieloo:2016bpk}%
  \BibitemOpen
  \bibfield  {author} {\bibinfo {author} {\bibfnamefont {A.}~\bibnamefont
  {Shafieloo}}, \bibinfo {author} {\bibfnamefont {D.~K.}\ \bibnamefont
  {Hazra}}, \bibinfo {author} {\bibfnamefont {V.}~\bibnamefont {Sahni}}, \ and\
  \bibinfo {author} {\bibfnamefont {A.~A.}\ \bibnamefont {Starobinsky}},\
  }\href@noop {} {\  (\bibinfo {year} {2016})},\ \Eprint
  {http://arxiv.org/abs/1610.05192} {arXiv:1610.05192 [astro-ph.CO]}
  \BibitemShut {NoStop}%
\bibitem [{\citenamefont {Aulakh}\ \emph {et~al.}(1998)\citenamefont {Aulakh},
  \citenamefont {Melfo},\ and\ \citenamefont {Senjanovic}}]{Aulakh:1998nn}%
  \BibitemOpen
  \bibfield  {author} {\bibinfo {author} {\bibfnamefont {C.~S.}\ \bibnamefont
  {Aulakh}}, \bibinfo {author} {\bibfnamefont {A.}~\bibnamefont {Melfo}}, \
  and\ \bibinfo {author} {\bibfnamefont {G.}~\bibnamefont {Senjanovic}},\
  }\href {\doibase 10.1103/PhysRevD.57.4174} {\bibfield  {journal} {\bibinfo
  {journal} {Phys. Rev.}\ }\textbf {\bibinfo {volume} {D57}},\ \bibinfo {pages}
  {4174} (\bibinfo {year} {1998})},\ \Eprint
  {http://arxiv.org/abs/hep-ph/9707256} {arXiv:hep-ph/9707256 [hep-ph]}
  \BibitemShut {NoStop}%
\bibitem [{\citenamefont {Duka}\ \emph {et~al.}(2000)\citenamefont {Duka},
  \citenamefont {Gluza},\ and\ \citenamefont {Zralek}}]{Duka:1999uc}%
  \BibitemOpen
  \bibfield  {author} {\bibinfo {author} {\bibfnamefont {P.}~\bibnamefont
  {Duka}}, \bibinfo {author} {\bibfnamefont {J.}~\bibnamefont {Gluza}}, \ and\
  \bibinfo {author} {\bibfnamefont {M.}~\bibnamefont {Zralek}},\ }\href
  {\doibase 10.1006/aphy.1999.5988} {\bibfield  {journal} {\bibinfo  {journal}
  {Annals Phys.}\ }\textbf {\bibinfo {volume} {280}},\ \bibinfo {pages} {336}
  (\bibinfo {year} {2000})},\ \Eprint {http://arxiv.org/abs/hep-ph/9910279}
  {arXiv:hep-ph/9910279 [hep-ph]} \BibitemShut {NoStop}%
\bibitem [{\citenamefont {Dobrescu}\ and\ \citenamefont
  {Liu}(2015)}]{Dobrescu:2015qna}%
  \BibitemOpen
  \bibfield  {author} {\bibinfo {author} {\bibfnamefont {B.~A.}\ \bibnamefont
  {Dobrescu}}\ and\ \bibinfo {author} {\bibfnamefont {Z.}~\bibnamefont {Liu}},\
  }\href {\doibase 10.1103/PhysRevLett.115.211802} {\bibfield  {journal}
  {\bibinfo  {journal} {Phys. Rev. Lett.}\ }\textbf {\bibinfo {volume} {115}},\
  \bibinfo {pages} {211802} (\bibinfo {year} {2015})},\ \Eprint
  {http://arxiv.org/abs/1506.06736} {arXiv:1506.06736 [hep-ph]} \BibitemShut
  {NoStop}%
\bibitem [{\citenamefont {Dobrescu}\ and\ \citenamefont
  {Fox}(2016)}]{Dobrescu:2015jvn}%
  \BibitemOpen
  \bibfield  {author} {\bibinfo {author} {\bibfnamefont {B.~A.}\ \bibnamefont
  {Dobrescu}}\ and\ \bibinfo {author} {\bibfnamefont {P.~J.}\ \bibnamefont
  {Fox}},\ }\href {\doibase 10.1007/JHEP05(2016)047} {\bibfield  {journal}
  {\bibinfo  {journal} {JHEP}\ }\textbf {\bibinfo {volume} {05}},\ \bibinfo
  {pages} {047} (\bibinfo {year} {2016})},\ \Eprint
  {http://arxiv.org/abs/1511.02148} {arXiv:1511.02148 [hep-ph]} \BibitemShut
  {NoStop}%
\bibitem [{\citenamefont {Ko}\ and\ \citenamefont {Nomura}(2016)}]{Ko:2015uma}%
  \BibitemOpen
  \bibfield  {author} {\bibinfo {author} {\bibfnamefont {P.}~\bibnamefont
  {Ko}}\ and\ \bibinfo {author} {\bibfnamefont {T.}~\bibnamefont {Nomura}},\
  }\href {\doibase 10.1016/j.physletb.2015.12.072} {\bibfield  {journal}
  {\bibinfo  {journal} {Phys. Lett.}\ }\textbf {\bibinfo {volume} {B753}},\
  \bibinfo {pages} {612} (\bibinfo {year} {2016})},\ \Eprint
  {http://arxiv.org/abs/1510.07872} {arXiv:1510.07872 [hep-ph]} \BibitemShut
  {NoStop}%
\bibitem [{\citenamefont {Bezrukov}\ \emph {et~al.}(2010)\citenamefont
  {Bezrukov}, \citenamefont {Hettmansperger},\ and\ \citenamefont
  {Lindner}}]{Bezrukov:2009th}%
  \BibitemOpen
  \bibfield  {author} {\bibinfo {author} {\bibfnamefont {F.}~\bibnamefont
  {Bezrukov}}, \bibinfo {author} {\bibfnamefont {H.}~\bibnamefont
  {Hettmansperger}}, \ and\ \bibinfo {author} {\bibfnamefont {M.}~\bibnamefont
  {Lindner}},\ }\href {\doibase 10.1103/PhysRevD.81.085032} {\bibfield
  {journal} {\bibinfo  {journal} {Phys. Rev.}\ }\textbf {\bibinfo {volume}
  {D81}},\ \bibinfo {pages} {085032} (\bibinfo {year} {2010})},\ \Eprint
  {http://arxiv.org/abs/0912.4415} {arXiv:0912.4415 [hep-ph]} \BibitemShut
  {NoStop}%
\bibitem [{\citenamefont {Esteves}\ \emph {et~al.}(2012)\citenamefont
  {Esteves}, \citenamefont {Romao}, \citenamefont {Hirsch}, \citenamefont
  {Porod}, \citenamefont {Staub},\ and\ \citenamefont
  {Vicente}}]{Esteves:2011gk}%
  \BibitemOpen
  \bibfield  {author} {\bibinfo {author} {\bibfnamefont {J.~N.}\ \bibnamefont
  {Esteves}}, \bibinfo {author} {\bibfnamefont {J.~C.}\ \bibnamefont {Romao}},
  \bibinfo {author} {\bibfnamefont {M.}~\bibnamefont {Hirsch}}, \bibinfo
  {author} {\bibfnamefont {W.}~\bibnamefont {Porod}}, \bibinfo {author}
  {\bibfnamefont {F.}~\bibnamefont {Staub}}, \ and\ \bibinfo {author}
  {\bibfnamefont {A.}~\bibnamefont {Vicente}},\ }\href {\doibase
  10.1007/JHEP01(2012)095} {\bibfield  {journal} {\bibinfo  {journal} {JHEP}\
  }\textbf {\bibinfo {volume} {01}},\ \bibinfo {pages} {095} (\bibinfo {year}
  {2012})},\ \Eprint {http://arxiv.org/abs/1109.6478} {arXiv:1109.6478
  [hep-ph]} \BibitemShut {NoStop}%
\bibitem [{\citenamefont {An}\ \emph {et~al.}(2012)\citenamefont {An},
  \citenamefont {Dev}, \citenamefont {Cai},\ and\ \citenamefont
  {Mohapatra}}]{An:2011uq}%
  \BibitemOpen
  \bibfield  {author} {\bibinfo {author} {\bibfnamefont {H.}~\bibnamefont
  {An}}, \bibinfo {author} {\bibfnamefont {P.~S.~B.}\ \bibnamefont {Dev}},
  \bibinfo {author} {\bibfnamefont {Y.}~\bibnamefont {Cai}}, \ and\ \bibinfo
  {author} {\bibfnamefont {R.~N.}\ \bibnamefont {Mohapatra}},\ }\href {\doibase
  10.1103/PhysRevLett.108.081806} {\bibfield  {journal} {\bibinfo  {journal}
  {Phys. Rev. Lett.}\ }\textbf {\bibinfo {volume} {108}},\ \bibinfo {pages}
  {081806} (\bibinfo {year} {2012})},\ \Eprint {http://arxiv.org/abs/1110.1366}
  {arXiv:1110.1366 [hep-ph]} \BibitemShut {NoStop}%
\bibitem [{\citenamefont {Nemevsek}\ \emph {et~al.}(2012)\citenamefont
  {Nemevsek}, \citenamefont {Senjanovic},\ and\ \citenamefont
  {Zhang}}]{Nemevsek:2012cd}%
  \BibitemOpen
  \bibfield  {author} {\bibinfo {author} {\bibfnamefont {M.}~\bibnamefont
  {Nemevsek}}, \bibinfo {author} {\bibfnamefont {G.}~\bibnamefont
  {Senjanovic}}, \ and\ \bibinfo {author} {\bibfnamefont {Y.}~\bibnamefont
  {Zhang}},\ }\href {\doibase 10.1088/1475-7516/2012/07/006} {\bibfield
  {journal} {\bibinfo  {journal} {JCAP}\ }\textbf {\bibinfo {volume} {1207}},\
  \bibinfo {pages} {006} (\bibinfo {year} {2012})},\ \Eprint
  {http://arxiv.org/abs/1205.0844} {arXiv:1205.0844 [hep-ph]} \BibitemShut
  {NoStop}%
\bibitem [{\citenamefont {Bhattacharya}\ \emph {et~al.}(2014)\citenamefont
  {Bhattacharya}, \citenamefont {Ma},\ and\ \citenamefont
  {Wegman}}]{Bhattacharya:2013nya}%
  \BibitemOpen
  \bibfield  {author} {\bibinfo {author} {\bibfnamefont {S.}~\bibnamefont
  {Bhattacharya}}, \bibinfo {author} {\bibfnamefont {E.}~\bibnamefont {Ma}}, \
  and\ \bibinfo {author} {\bibfnamefont {D.}~\bibnamefont {Wegman}},\ }\href
  {\doibase 10.1140/epjc/s10052-014-2902-7} {\bibfield  {journal} {\bibinfo
  {journal} {Eur. Phys. J.}\ }\textbf {\bibinfo {volume} {C74}},\ \bibinfo
  {pages} {2902} (\bibinfo {year} {2014})},\ \Eprint
  {http://arxiv.org/abs/1308.4177} {arXiv:1308.4177 [hep-ph]} \BibitemShut
  {NoStop}%
\bibitem [{\citenamefont {Heeck}\ and\ \citenamefont
  {Patra}(2015)}]{Heeck:2015qra}%
  \BibitemOpen
  \bibfield  {author} {\bibinfo {author} {\bibfnamefont {J.}~\bibnamefont
  {Heeck}}\ and\ \bibinfo {author} {\bibfnamefont {S.}~\bibnamefont {Patra}},\
  }\href {\doibase 10.1103/PhysRevLett.115.121804} {\bibfield  {journal}
  {\bibinfo  {journal} {Phys. Rev. Lett.}\ }\textbf {\bibinfo {volume} {115}},\
  \bibinfo {pages} {121804} (\bibinfo {year} {2015})},\ \Eprint
  {http://arxiv.org/abs/1507.01584} {arXiv:1507.01584 [hep-ph]} \BibitemShut
  {NoStop}%
\bibitem [{\citenamefont {Garcia-Cely}\ and\ \citenamefont
  {Heeck}(2015)}]{Garcia-Cely:2015quu}%
  \BibitemOpen
  \bibfield  {author} {\bibinfo {author} {\bibfnamefont {C.}~\bibnamefont
  {Garcia-Cely}}\ and\ \bibinfo {author} {\bibfnamefont {J.}~\bibnamefont
  {Heeck}},\ }\href {\doibase 10.1088/1475-7516/2016/03/021} {\bibfield
  {journal} {\bibinfo  {journal} {JCAP}\ }\textbf {\bibinfo {volume} {1603}},\
  \bibinfo {pages} {021} (\bibinfo {year} {2015})},\ \Eprint
  {http://arxiv.org/abs/1512.03332} {arXiv:1512.03332 [hep-ph]} \BibitemShut
  {NoStop}%
\bibitem [{\citenamefont {Berlin}\ \emph {et~al.}(2016)\citenamefont {Berlin},
  \citenamefont {Fox}, \citenamefont {Hooper},\ and\ \citenamefont
  {Mohlabeng}}]{Berlin:2016eem}%
  \BibitemOpen
  \bibfield  {author} {\bibinfo {author} {\bibfnamefont {A.}~\bibnamefont
  {Berlin}}, \bibinfo {author} {\bibfnamefont {P.~J.}\ \bibnamefont {Fox}},
  \bibinfo {author} {\bibfnamefont {D.}~\bibnamefont {Hooper}}, \ and\ \bibinfo
  {author} {\bibfnamefont {G.}~\bibnamefont {Mohlabeng}},\ }\href {\doibase
  10.1088/1475-7516/2016/06/016} {\bibfield  {journal} {\bibinfo  {journal}
  {JCAP}\ }\textbf {\bibinfo {volume} {1606}},\ \bibinfo {pages} {016}
  (\bibinfo {year} {2016})},\ \Eprint {http://arxiv.org/abs/1604.06100}
  {arXiv:1604.06100 [hep-ph]} \BibitemShut {NoStop}%
\bibitem [{\citenamefont {Coleman}(1977)}]{Coleman:1977py}%
  \BibitemOpen
  \bibfield  {author} {\bibinfo {author} {\bibfnamefont {S.~R.}\ \bibnamefont
  {Coleman}},\ }\href {\doibase 10.1103/PhysRevD.15.2929,
  10.1103/PhysRevD.16.1248} {\bibfield  {journal} {\bibinfo  {journal} {Phys.
  Rev.}\ }\textbf {\bibinfo {volume} {D15}},\ \bibinfo {pages} {2929} (\bibinfo
  {year} {1977})},\ \bibinfo {note} {[Erratum: Phys.
  Rev.D16,1248(1977)]}\BibitemShut {NoStop}%
\bibitem [{\citenamefont {Weinberg}(2013)}]{Weinberg:1996kr}%
  \BibitemOpen
  \bibfield  {author} {\bibinfo {author} {\bibfnamefont {S.}~\bibnamefont
  {Weinberg}},\ }\href@noop {} {\emph {\bibinfo {title} {{The quantum theory of
  fields. Vol. 2: Modern applications}}}}\ (\bibinfo  {publisher} {Cambridge
  University Press},\ \bibinfo {year} {2013})\BibitemShut {NoStop}%
\bibitem [{\citenamefont {Coleman}\ \emph {et~al.}(1978)\citenamefont
  {Coleman}, \citenamefont {Glaser},\ and\ \citenamefont
  {Martin}}]{Coleman:1977th}%
  \BibitemOpen
  \bibfield  {author} {\bibinfo {author} {\bibfnamefont {S.~R.}\ \bibnamefont
  {Coleman}}, \bibinfo {author} {\bibfnamefont {V.}~\bibnamefont {Glaser}}, \
  and\ \bibinfo {author} {\bibfnamefont {A.}~\bibnamefont {Martin}},\ }\href
  {\doibase 10.1007/BF01609421} {\bibfield  {journal} {\bibinfo  {journal}
  {Commun. Math. Phys.}\ }\textbf {\bibinfo {volume} {58}},\ \bibinfo {pages}
  {211} (\bibinfo {year} {1978})}\BibitemShut {NoStop}%
\bibitem [{\citenamefont {Coleman}\ and\ \citenamefont
  {De~Luccia}(1980)}]{Coleman:1980aw}%
  \BibitemOpen
  \bibfield  {author} {\bibinfo {author} {\bibfnamefont {S.~R.}\ \bibnamefont
  {Coleman}}\ and\ \bibinfo {author} {\bibfnamefont {F.}~\bibnamefont
  {De~Luccia}},\ }\href {\doibase 10.1103/PhysRevD.21.3305} {\bibfield
  {journal} {\bibinfo  {journal} {Phys. Rev.}\ }\textbf {\bibinfo {volume}
  {D21}},\ \bibinfo {pages} {3305} (\bibinfo {year} {1980})}\BibitemShut
  {NoStop}%
\bibitem [{\citenamefont {Olive}\ \emph {et~al.}(2014)\citenamefont {Olive}
  \emph {et~al.}}]{Agashe:2014kda}%
  \BibitemOpen
  \bibfield  {author} {\bibinfo {author} {\bibfnamefont {K.~A.}\ \bibnamefont
  {Olive}} \emph {et~al.} (\bibinfo {collaboration} {Particle Data Group}),\
  }\href {\doibase 10.1088/1674-1137/38/9/090001} {\bibfield  {journal}
  {\bibinfo  {journal} {Chin. Phys.}\ }\textbf {\bibinfo {volume} {C38}},\
  \bibinfo {pages} {090001} (\bibinfo {year} {2014})}\BibitemShut {NoStop}%
\end{thebibliography}%

\end{document}